\documentclass[twocolumn,amssymb,nobibnotes,aps, reprint,letterpaper,pre/printnumbers,amsmath,prb,
showpacs,floatfix]{revtex4-2}
\usepackage{graphicx,xcolor}% Include figure files [dvips]
\usepackage{dcolumn}% Align table columns on decimal point
\usepackage{bm}% bold math
\usepackage{graphics}
\usepackage{float}
\usepackage{epsfig}
\usepackage{multirow}
\usepackage{amsmath}
\usepackage{tabularx}
\usepackage{mathrsfs}
\usepackage[pass,paperwidth=8.5in,paperheight=11in]{geometry}
\usepackage[linktocpage=true]{hyperref}
\usepackage{hyperref}
\usepackage{comment}
\usepackage{soul}
\usepackage{notes2bib}
\hypersetup{
  pdfnewwindow=true, 
  colorlinks=true,
  linkcolor=blue, 
  anchorcolor=blue,
  citecolor=blue, 
  filecolor=blue,
  menucolor=blue, 
  urlcolor=blue}

\usepackage[normalem]{ulem}
\usepackage{xcolor}
\usepackage{orcidlink}

\begin{document}
 
\date{\today}

\title{Kagome KMn$_3$Sb$_5$  metal: Magnetism, lattice dynamics, and anomalous Hall conductivity}

\author{Sobhit Singh\,\orcidlink{0000-0002-5292-4235}}
\email{s.singh@rochester.edu}
\affiliation{Department of Mechanical Engineering, University of Rochester, Rochester, New York 14627, USA}

\author{A. C. Garcia-Castro\,\orcidlink{0000-0003-3379-4495}}
\email{acgarcia@uis.edu.co}
\affiliation{School of Physics, Universidad Industrial de Santander, Carrera 27 Calle 09, 680002, Bucaramanga, Colombia}

%%%%%%%%%%%%%%%%%%%%%%%%%%%%%%%%%%%%%%%%%%%%%%%%
\begin{abstract} 
Kagome metals are reported to exhibit remarkable properties, including superconductivity, charge density wave order, and a large anomalous Hall conductivity, which facilitate the implementation of spintronic devices.
In this work, we study a novel kagome metal based on Mn magnetic sites in a KMn$_3$Sb$_5$ stoichiometry. 
By means of first-principles density functional theory calculations, we demonstrate that the studied compound is dynamically stable, locking the ferromagnetic order as the ground state  configuration, thus preventing the charge-density-wave state as reported in its vanadium-based counterpart KV$_3$Sb$_5$. 
Our calculations predict that KMn$_3$Sb$_5$ exhibits an out-of-plane (001) ferromagnetic response as the ground state, allowing for 
the emergence of topologically protected Weyl nodes near the Fermi level and nonzero anomalous Hall conductivity ($\sigma_{ij}$) in this centrosymmetric system.
We obtain a tangible $\sigma_{xy} = 314$ S$\cdot$cm$^{-1}$ component, which is comparable to that of other kagome metals. 
Finally, we explore the effect of the on-site Coulomb repulsion ($+U$) on the structural and electronic properties and find that, although the lattice parameters and $\sigma_{xy}$ moderately vary with increasing $+U$, KMn$_3$Sb$_5$ stands as an ideal stable ferromagnetic kagome metal with a large anomalous Hall conductivity response.\\
\\
DOI:

\end{abstract}

\maketitle

%%%%%%%%%%%%%%%%%%%%%%%%%%%%%%%%%%%%%%%%%%%%%%%%
\section{Introduction}
Kagome lattices \cite{doi:10.1063/1.1564329}, such as those observed in FeSn \cite{FeSn-NatMat-2020}, Co$_3$Sn$_2$S$_2$ \cite{Co3Sn2S2-kagome-NatPhys-2018}, ScV$_6$Sn$_6$ \cite{PhysRevLett.129.216402}, and the ones present in Mn$_{3}B$N ($B$ = Ni, Ga, Pt, Pd, and Sn) and V$_3$AuN antiperovskites \cite{Wang2019, Garcia-Castro2020, Garcia-Castro2019,PhysRevB.106.195113,FLOREZGOMEZ2022169813,PhysRevMaterials.6.125003}, exhibit remarkable electronic, phononic, topological, and magnetic entangled properties, mainly owing to their particular star-shaped hexagonal symmetry \cite{NatMat-2020-1}.
In such symmetry, triangularly-coordinated magnetically active cations present substantial magnetic and electronic frustration that leads, for example, to charge-density waves (CDW) phases \cite{AV3Sb5-NPJ-2022}, superconductivity \cite{Kagome-nature-2023} and chiral noncollinear magnetic orderings \cite{NatMat-2020-1}. 
Belonging to the kagome materials, the $A$V$_3$Sb$_5$ ($A$ = K, Rb, and Cs) family has recently attracted tremendous attention due to the richness of their charge-density wave states, giant anomalous Hall response, frustrated electronic structure, and associated superconductivity \cite{PhysRevMaterials.3.094407,NatComm-2023-1,SciRep-2022-1,PhysRevB.107.205131,ritz2023superconductivity}. %\ss{[cite BIROL]}.
The reported CDW phase in $A$V$_3$Sb$_5$ metals is originated by the electronic frustration in the in-plane vanadium sites, due to the unbalanced charge based on the expected nominal charges, $i.e.$, K:$1+$, V:(5-$\delta$)$+$ and Sb:$3-$, to enforce the charge neutrality with K$^{1+}$V$_3^{(5-\delta)+}$Sb$_5^{3-}$ \cite{PhysRevMaterials.3.094407}.
Moreover, ferromagnetic and noncollinear antiferromagnetic states can be expected to stabilize in this family of materials, leading to tangible topological features related to Dirac and Weyl fermions in the vicinity of the Fermi level, resulting in non-vanishing Berry curvature induced observables such as anomalous Hall conductivity (AHC) and nonlinear Hall effects \cite{PhysRevLett.112.017205, AHC-AFM-NatRev-2022, Souza2007, DV2006, SS2020}.
Importantly, the electrical manipulation of Berry curvature-induced anomalous Hall effect at room temperature offers unimaginable capabilities in future spintronic devices \cite{Liu2018, Tsai2020, Qin2019, doi:10.1063/5.0101981}. In ferromagnetic kagome metals, reversible magnetization can be used as an external parameter to control and switch the AHC response. 
Therefore, the search for novel ferromagnetic kagome materials, belonging to the $A$$M$$_3$Sb$_5$ (with $A$ = K, Rb, Cs. $M$ = V and Mn) stoichiometry, might lead to the discovery of novel quantum phases driven by nontrivial electronic, topological, magnetic, and superconducting properties. 

In this study, we employ first-principles density functional theory (DFT) calculations to investigate the structural, vibrational, magnetic, and topological electronic properties of the kagome metal KMn$_3$Sb$_5$, which was recently predicted by Jiang \emph{et al.}~\cite{Jiang_2022}. 
Our results reveal that, despite the inherent magnetic frustration in the kagome Mn plane, KMn$_3$Sb$_5$ exhibits a stable ferromagnetic ground state, with the magnetic easy axis oriented along the (001) direction. 
This ferromagnetic ground state breaks time-reversal symmetry and allows the emergence of topological Weyl nodes in this centrosymmetric crystal system, resulting in large concentrations of Berry curvature near the Weyl nodes. 
As a result, KMn$_3$Sb$_5$ exhibits a substantial anomalous Hall conductivity response ($\sigma_{xy}$ = 314 S$\cdot$cm$^{-1}$).
Incorporating the on-site Coulomb parameter $+U$ in our DFT calculations moderately affects the optimized lattice parameters, magnetic moments, phonon frequencies, and $\sigma_{xy}$ values, but no unstable phonon modes are observed within the studied range of $+U$ parameter. 
Our findings provide valuable insights for experimentalists in synthesizing and confirming the predicted properties of the ferromagnetic kagome metal KMn$_3$Sb$_5$.

%%%%%%%%%%%%%%%%
\begin{figure*}[]
 \centering
 \includegraphics[width=18.0cm,keepaspectratio=true]{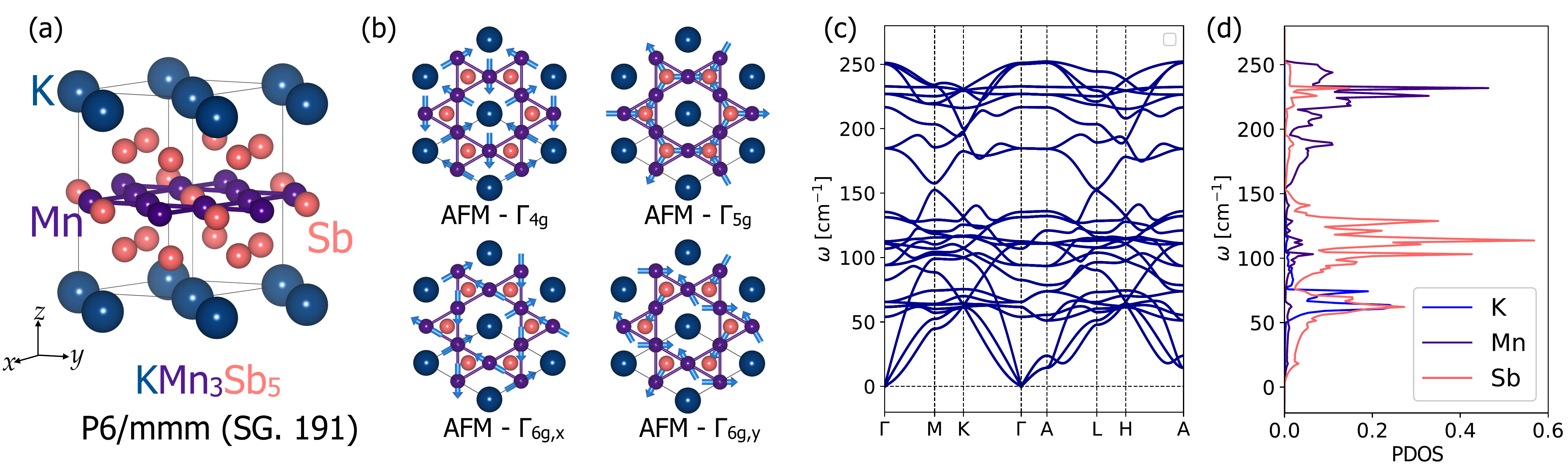}
 \caption{(Color online) (a) KMn$_3$Sb$_5$ \emph{P6/mmm} (SG.\,191) hexagonal structure obtained for the ferromagnetic, (001) FM, ground state. In this structure, the K, Mn, and Sb sites are drawn in dark blue, violet, and pink colors, respectively. (b) Chiral noncollinear antiferromagnetic orderings allowed in KMn$_3$Sb$_5$. Here, the $\Gamma_{4g}$ and $\Gamma_{5g}$ AFM orders hold $+1$ magnetic chirality whereas the $\Gamma_{6g,x}$ and $\Gamma_{6g,y}$ AFM orders inherit $-1$ magnetic chirality. (c) Phonon dispersion calculated for the ground state FM order (DFT-PBEsol). 
 (d) Atom-projected phonon density of states (PDOS).}
 \label{F1}
\end{figure*} 
%%%%%%%%%%%%%%%%

This paper is organized as follows: In Section~\ref{secII}, we provide details of the computational and theoretical methods used in this study. In Section~\ref{secIII}, we present the results, starting with a discussion of the crystal structure and its vibrational and dynamical stability. We then analyze different magnetic configurations and explore the electronic structure, including intriguing features associated with Berry curvature and anomalous Hall conductivity. Furthermore, we examine the role of electron-correlation effects on the properties of KMn$_3$Sb$_5$. Finally, in Section~\ref{secIV}, we draw our conclusions and discuss future perspectives.

%%%%%%%%%%%%%%%%%%%%%%%%%%%%%%%%%%%%%%%%%%%%%%%%
\section{Computational Details} \label{secII}
We performed first-principles DFT calculations \cite{PhysRev.136.B864,PhysRev.140.A1133} using the projected-augmented wave (PAW) \cite{Blochl1994} method as implemented in the \textsc{vasp} code (version 5.4.4) \cite{Kresse1996,Kresse1999}. 
%The projected-augmented waves, PAW \cite{Blochl1994}, approach representing the valence and core electrons was used. 
The valence electron configurations considered in the PAW pseudopotentials are as follows: K: (3$p^6$4$s^1$, version 02Aug2007), Mn: (3$p^6$3$d^5$4$s^2$, version 02Aug2007), and Sb: (5$s^2$5$p^3$, version 06Sep2000). 
The exchange-correlation functional was computed using the generalized-gradient approximation as parameterized by Perdew-Burke-Ernzerhof for solids (GGA-PBEsol) \cite{Perdew2008} and 
the on-site correlation effects for Mn-3$d$ electrons were corrected using the rotationally-invariant Liechtenstein (DFT$+U$) formalism \cite{Liechtenstein1995}.
We used $+U$ values ranging from 0.0 to 3.0\,eV, which were optimized to explore the behavior of the lattice parameters, magnetic moment, and anomalous Hall conductivity.
The reciprocal space was sampled using a $\Gamma$-centered Monkhorst-Pack $k$-mesh \cite{PhysRevB.13.5188} of size 11$\times$11$\times$9, and a kinetic energy cutoff of 600\,eV was used for the plane wave basis set. These values resulted in the convergence of residual forces and total energy to better than 0.001\,eV$\cdot$\r{A}$^{-1}$ and 0.1\,meV. 
Spin-orbit coupling (SOC) was included to consider noncollinear magnetic configurations \cite{Hobbs2000}. Phonon calculations were performed within the finite-differences approach \cite{PhysRevLett.48.406, PhysRevB.34.5065} and post-processed using the \textsc{Phonopy} code \cite{phonopy}.
To compute the anomalous Hall conductivity and Berry curvature, we utilized the Wannier functions methodology, for which the wannierization was performed using the \textsc{Wannier90} code \cite{MOSTOFI20142309, Pizzi_2020} and post-processed with the \textsc{WannierBerri} package \cite{wannierberri}. 
For the wannierization process, $s$ and $p$ orbitals were considered for K and Sb atoms, while $s$, $p$, and $d$ were considered for Mn atoms. 
To plot the Berry curvature projected on the Fermi surface, we utilized the \textsc{FermiSurfer} software \cite{KAWAMURA2019197}.
Crystal structures were visualized using the \textsc{vesta} software \cite{vesta}, and electronic structure data were post-processed using the \textsc{PyProcar} \cite{HERATH2020107080} software.

%%%%%%%%%%%%%%%%%%%%%%%%%%%%%%%%%%%%%%%%%%%%%%%%
\section{Results and Discussion}\label{secIII}

%%%%%%%%%%%%%%%%%%%%%%%%%%%%%%%%%%%%%%%%%%%%%%%%
\subsection{Structural and Magnetic Configurations:}

Similar to $A$V$_3$Sb$_5$, KMn$_3$Sb$_5$ adopts the $P6/mmm$ (SG.\,191) phase in its primitive unit cell. It consists of an Mn kagome lattice situated in the (0,\,0,\,1/2) plane, with several Sb sites embedded into the central kagome hexagon, as shown in Fig.~\ref{F1}(a). 
Additionally, the Mn kagome lattice is packed with graphite-like Sb layers in positions close to (0,\,0,\,1/4) and (0,\,0,\,3/4) atomic planes. 
Lastly, the hexagonally coordinated K atoms are located at the (0,\,0,\,0) sites.
The DFT-optimized lattice parameters of KMn$_3$Sb$_5$ are provided below in Table~\ref{tab:1}.

As a first step in the magnetic analysis, we define all the symmetry-allowed noncollinear magnetic states that can be associated with a {\bf q} propagation vector of (0,\,0,\,0). 
We find that, similar to the case of antiperovskites~\cite{PhysRevMaterials.6.125003}, four $xy$ in-plane noncollinear chiral antiferromagnetic (AFM) orderings are allowed, as shown in Fig.~\ref{F1}(b), along with a $z$-axis out-of-plane ferromagnetic (FM) ordering.

We then calculated the total energies (PBEsol) for all the considered noncollinear magnetic orderings and obtained 
$E_{\Gamma_{4g}}$ = $-$54.0587 eV$\cdot$f.u.$^{-1}$ and $E_{\Gamma_{5g}}$ = $-$54.0614 eV$\cdot$f.u.$^{-1}$ for the chiral $+1$, and 
$E_{\Gamma_{6g,x}}$ = $-$54.0453 eV$\cdot$f.u.$^{-1}$ and  $E_{\Gamma_{6g,y}}$ = $-$54.0452 eV$\cdot$f.u.$^{-1}$ for the chiral $-1$ orderings~\cite{doi:10.1143/JPSJ.53.4138}.
See note~\footnote{The magnetic vector chirality is defined as \textbf{$\kappa$} = $\frac{2}{3 \sqrt{3}}\sum_{i,j}$(\textbf{S$_i$}$\times$\textbf{S$_j$}) = $\frac{2}{3 \sqrt{3}}$(\textbf{S$_1$}$\times$\textbf{S$_2$}+\textbf{S$_2$}$\times$\textbf{S$_3$}+\textbf{S$_3$}$\times$\textbf{S$_1$}) for the kagome lattice where the $i$ and $j$ index run over the magnetic moments in the unit cell} for more details regarding magnetic chirality. 
Interestingly, we find that the (001) FM state, with a total energy of $E_{FM}$ = $-$54.4928 eV$\cdot$f.u.$^{-1}$ is the lowest energy magnetic ground state with an average difference energy around 500 meV$\cdot$f.u.$^{-1}$. 
This is in contrast with the observed ground state magnetic configuration in other similar materials such as the antiperovskites. In antiperovskites, the 3D kagome symmetry resolves magnetic frustration by stabilizing chiral noncollinear antiferromagnetic states instead of ferromagnetic states.

The ground state FM ordering in KMn$_3$Sb$_5$ can be explained in terms of the out-of-plane broken magnetic frustration which tends to lower the total energy compared to the strong frustrations present in the kagome lattice of antiperovskites. 
In antiperovskites, the kagome lattices are formed, by symmetry, in the complete ensemble of [111] family of planes, leading to the stabilization of chiral noncollinear antiferromagnetic states due to the 3D kagome symmetry. However, in KMn$_3$Sb$_5$, the out-of-plane magnetic frustration breaks this symmetry and favors the ferromagnetic ordering as the ground state magnetic configuration.

To gain a better understanding of the magnetic easy axis, we investigate the preferred direction in which the magnetic moments tend to align in the kagome plane of Mn. Specifically, we examine whether the moments lie in the $xy$-plane (within the plane) or if they are oriented perpendicular to it (along the $z$-axis).
Our calculations (PBEsol) reveal that $E_{xy-plane}$ = $-$54.4920 eV$\cdot$f.u.$^{-1}$ whereas 
$E_{z-axis}$ = $-$54.4928 eV$\cdot$f.u.$^{-1}$, thus 
implying $z$-axis as the easy axis for the ground state FM order in KMn$_3$Sb$_5$. 

From this point on, all the reported calculations and analyses are for the (001) ferromagnetic ground state of KMn$_3$Sb$_5$. 
The DFT (PBEsol) optimized lattice parameters in this magnetic ground state are $a=b$ = 5.337 \r{A} and $c$ = 9.028 \r{A}. 
With increasing $+U$ value (PBEsol$+U$) from 0 to 3\,eV on Mn-$3d$ orbitals, primitive cell moderately expands within the $a$-$b$ plane while shrinking along the $c$ axis (see Table~\ref{tab:1}) by approximately $\pm5\%$. 

% a-b lattice parameters:
% 5.602-5.337 = 0.265 Å 
% 0.265/5.337 => 4.97%  Expansion wth increasing U

% c lattice parameter (U=0 to U = 3 eV)
% 9.028-8.606 = 0.422
% 0.422/9.028 => 4.7% contraction with increasing U

%%%%%%%%%%%%%%%%%%%%%%%%%%%%%%%%%%%%%%%%%%%%%%%%
\subsection{Lattice Dynamics:}
To test the dynamical stability of the ground state noncollinear (001) FM order in KMn$_3$Sb$_5$, we calculated the full phonon spectrum along the high-symmetry directions of the Brillouin zone, as shown in Fig.~\ref{F1}(c).
We observe a fully stable vibrational landscape, with no unstable modes at imaginary frequencies appearing along the entire Brillouin zone.
The absence of unstable phonons suggests the suppression of the charge-density-wave phenomenon induced by the unstable phonon modes, as reported in other kagome systems such as $A$V$_3$Sb$_5$ \cite{NatComm-2023-1,PhysRevB.107.205131,ritz2023superconductivity}.  
Thus, it is important to note that the kagome compound KMn$_3$Sb$_5$ is added to the set of materials with this symmetry, presenting potential opportunities for exhibiting a tangible ferromagnetic response.
According to group theory, the irreducible representation of all allowed vibrational modes for KMn$_3$Sb$_5$ ($P6/mmm$) at Brillouin zone center is defined as follows: 

\begin{equation}\label{eq1}
\begin{aligned}
\Gamma\textsubscript{vib} = \emph{A}\textsubscript{1g} \oplus
4\emph{A}\textsubscript{2u} \oplus  \emph{B}\textsubscript{1g} \oplus 
\emph{B}\textsubscript{1u} \oplus 2\emph{B}\textsubscript{2u}  \oplus
 2\emph{E}\textsubscript{2u} \\ \oplus \, \emph{E}\textsubscript{2g} \oplus 5\emph{E}\textsubscript{1u} \oplus 
 \emph{E}\textsubscript{1g}. 
\end{aligned}
\end{equation}

Out of the total 27 allowed phonon modes (9 atoms per cell), the three acoustic modes are 
$\Gamma\textsubscript{acoustic}$ = $\emph{A}\textsubscript{2u} \oplus  \emph{E}\textsubscript{1u}$, and 
24 optic modes are 
$\Gamma\textsubscript{optic}$ = 
$ \emph{A}\textsubscript{1g} \oplus 
3\emph{A}\textsubscript{2u} \oplus 
\emph{B}\textsubscript{1g}  \oplus 
\emph{B}\textsubscript{1u}  \oplus 
2\emph{B}\textsubscript{2u}  \oplus 
2\emph{E}\textsubscript{2u}  \oplus 
\emph{E}\textsubscript{2g}  \oplus 
4\emph{E}\textsubscript{1u}  \oplus 
\emph{E}\textsubscript{1g} $. 
Here, 
 $\emph{A}\textsubscript{1g}$, 
 $\emph{E}\textsubscript{2g}$, and $\emph{E}\textsubscript{1g}$ modes are Raman active, whereas 
 $\emph{A}\textsubscript{2u}$ and $\emph{E}\textsubscript{1u}$ modes are infrared (IR) active. All other modes are silent. 
%[\ss{Camilo, please double check the above details with Bilbao and correct these as needed. Reply: I did double-check and this is the Irrep}]

Table~\ref{tab:modes} presents the zone center phonon frequencies calculated using PBEsol ($U$ = 0.0 eV) for the ground state (001) FM ordering in KMn$_3$Sb$_5$. As $U$ is increased, the phonon frequencies vary, but no phonon instability was observed within the range of studied $U$ values (see for example the full-phonon dispersion and the phonon DOS at $U$ = 3.0 eV in Fig. \ref{F2a} in Appendix \ref{secA}).
To fully provide the behavior of the active Raman and IR frequencies, as a function of $U$, we present the $A_{1g}$, $A_{2g}$, and $E_{1g}$ modes in Fig. \ref{F2a-2}(a), (in Appendix \ref{secA}). The $A_{2u}$ and $E_{1u}$ modes are presented in Fig. \ref{F2a-2}(b). In the overall trend, we can appreciate a softening of the modes when the $U$ value is increased. 

%%%%%%%%%%%%%%%%
\begin{table}[b]
\centering
\caption{DFT-PBEsol calculated zone center phonon frequencies for the (001) FM ground state of KMn$_3$Sb$_5$. Data for different values of $+U$ parameter are provided in Appendix \ref{secA}.}
\begin{tabular}{c  c | c  c }
\hline
\hline
Mode  &  $\omega$ (cm$^{-1}$) &  Mode &   $\omega$ (cm$^{-1}$)  \rule[-1ex]{0pt}{3.5ex} \\
\hline
$A_{1g}$ &  106.8 &  $A_{2u}$ & 53.9, 82.5, 232.8   \rule[-1ex]{0pt}{3.5ex}  \\
$B_{1g}$ &  135.3 &  $B_{1u}$ & 112.5, 251.1    \rule[-1ex]{0pt}{3.5ex}  \\
$E_{2g}$ &  131.3 &  $B_{2u}$ & 216.7   \rule[-1ex]{0pt}{3.5ex}  \\
$E_{1g}$ &  65.5 &  $E_{2u}$& 110.7, 227.0  \rule[-1ex]{0pt}{3.5ex}  \\
---            &  ---  &  $E_{1u}$& 61.6, 94.1, 184.7, 250.2  \rule[-1ex]{0pt}{3.5ex}  \\
\hline
\hline
\end{tabular}
\label{tab:modes}
\end{table}
%%%%%%%%%%%%%%%%

Interestingly, we observe that, based on the atomically-project phononic density-of-states, in Fig. \ref{F1}(d), the Mn sites contribute strongly at the high-frequency modes, in between 150 to 250 cm$^{-1}$. 
On the contrary, the strongest Sb contribution can be noted between 0 cm$^{-1}$ to 150 cm$^{-1}$. Finally, the K atom's contribution to the lattice dynamics is strongly localized around 60 cm$^{-1}$.

%%%%%%%%%%%%%%%%%%%%%%%%%%%%%%%%%%%%%%%%%%%%%%%%
\subsection{Electronic Structure:}

%%%%%%%%%%%%%%%%
\begin{figure}[t]
 \centering
 \includegraphics[width=8.6cm,keepaspectratio=true]{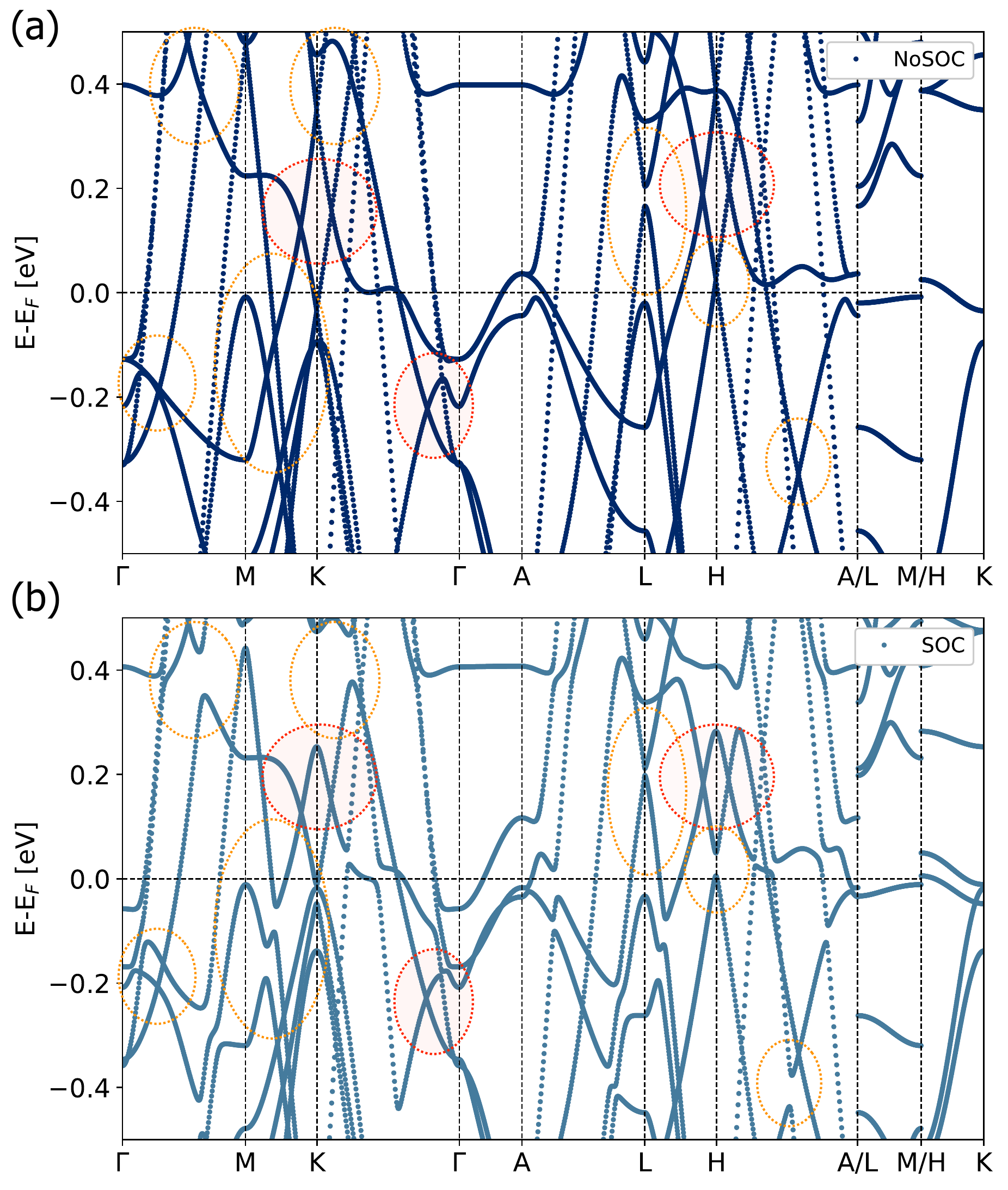}
 \caption{(Color online) Electronic band structure computed in the ferromagnetic ground state for (a) without SOC and (b) with SOC cases. For the SOC case, a (001) noncollinear ferromagnetic configuration was used. Here, in orange ovals, are marked the gapped nodes along the high-symmetry $k$ directions whereas, in red ovals, are shown the symmetry-protected Weyl nodes. The latter are located along the $M$--$K$, and $\Gamma$--$K$ paths at the equivalent points in the (0,\,0,\,1/2) plane.}
 \label{F2}
\end{figure} 
%%%%%%%%%%%%%%%%

Moving forward, in Fig.~\ref{F2} we present the PBEsol calculated electronic bands obtained with and without considering SOC effects. 
Here, it is worth recalling that noncollinear magnetic moments of Mn are aligned along the $z$-axis.
As expected, our band structure calculations reveal metallic features. We note that the inclusion of $+U$ in our DFT calculations does not result in the opening of any bandgap; it only moderately modifies the details of the bands near the Fermi level (see, for example, the electronic band dispersion obtained for $U$ = 3.0 eV Fig. \ref{F3a} in Appendix \ref{secB}.).

At first glance, in the absence of SOC, we notice multiple band crossings near the Fermi level that might be associated with potential topological nodes, in Fig.~\ref{F2}(a). 
Some of these band crossings become gapped once the SOC effects are included, leading to the emergence of a multitude of topologically protected Weyl nodes in the vicinity of these gapped band crossings and away from the high-symmetry k-path considered in Fig.~\ref{F2}(b)~\cite{Gapped-nodes-Nature-2018, SS_PRB2016, Karki2022}. 
Additionally, several crystal symmetry-protected nodes occur along the high-symmetry directions of the Brillouin zone, as expected for the kagome symmetry~\cite{Co3Sn2S2-kagome-NatPhys-2018,PhysRevB.101.115106}.
These crystal symmetry-protected nodes are located, for example, along the $M$--$K$ and $\Gamma$--$K$ k-paths at energies close to 150\,meV above the Fermi level. 
There are symmetry-protected nodes also located at the $L$--$H$ and $H$--$A$ in the (0,\,0,\,1/2) $k$-plane.
All these Weyl nodes serve as sources and sinks of Berry curvature in the momentum space yielding a large AHC response in the FM kagome metal KMn$_3$Sb$_5$, as we discuss below. 
As expected, the spin-projection of the Weyl nodes is reversed, as can be appreciated from the spin-polarized electronic band structure shown in Fig. \ref{F1a} at Appendix \ref{secC}. Here, only the $S_z$ component is observed in the vicinity of the Fermi energy, which is in agreement with the underlying magnetic structure.

%%%%%%%%%%%%%%%%%%%%%%%%%%%%%%%%%%%%%%%%%%%%%%%%
\subsection{Anomalous Hall Conductivity:}

By analyzing the symmetry-allowed properties and electronic structure, we notice that the ferromagnetic kagome metal KMn$_3$Sb$_5$ may exhibit a tangible anomalous Hall conductivity response. Consequently, in the \emph{P6/mm'm'} magnetic space group (MSG.\,191.240), the AHC tensor ($\sigma_{ij}$) is extracted and presented in Eq.~\ref{ahc:tensor}. 
Most of the AHC tensor elements are forced to vanish due to symmetry, except for $\sigma_{xy}$ and $\sigma_{yx}$ which are allowed to have nonzero values. However, it is important to note that $\sigma_{yx}$ = $-\sigma_{xy}$. Thus, the nonzero AHC components lie within the kagome plane and perpendicular to the orientation of Mn magnetic moments breaking the time-reversal symmetry. 

\begin{align}\label{ahc:tensor}
\sigma_{6/mm'm'}=
\begin{pmatrix}
0 & \sigma_{xy} & 0\\
-\sigma_{xy} & 0 & 0 \\
0 & 0 & 0
\end{pmatrix}
\end{align}

\begin{figure}[t]
 \centering
 \includegraphics[width=8.2cm,keepaspectratio=true]{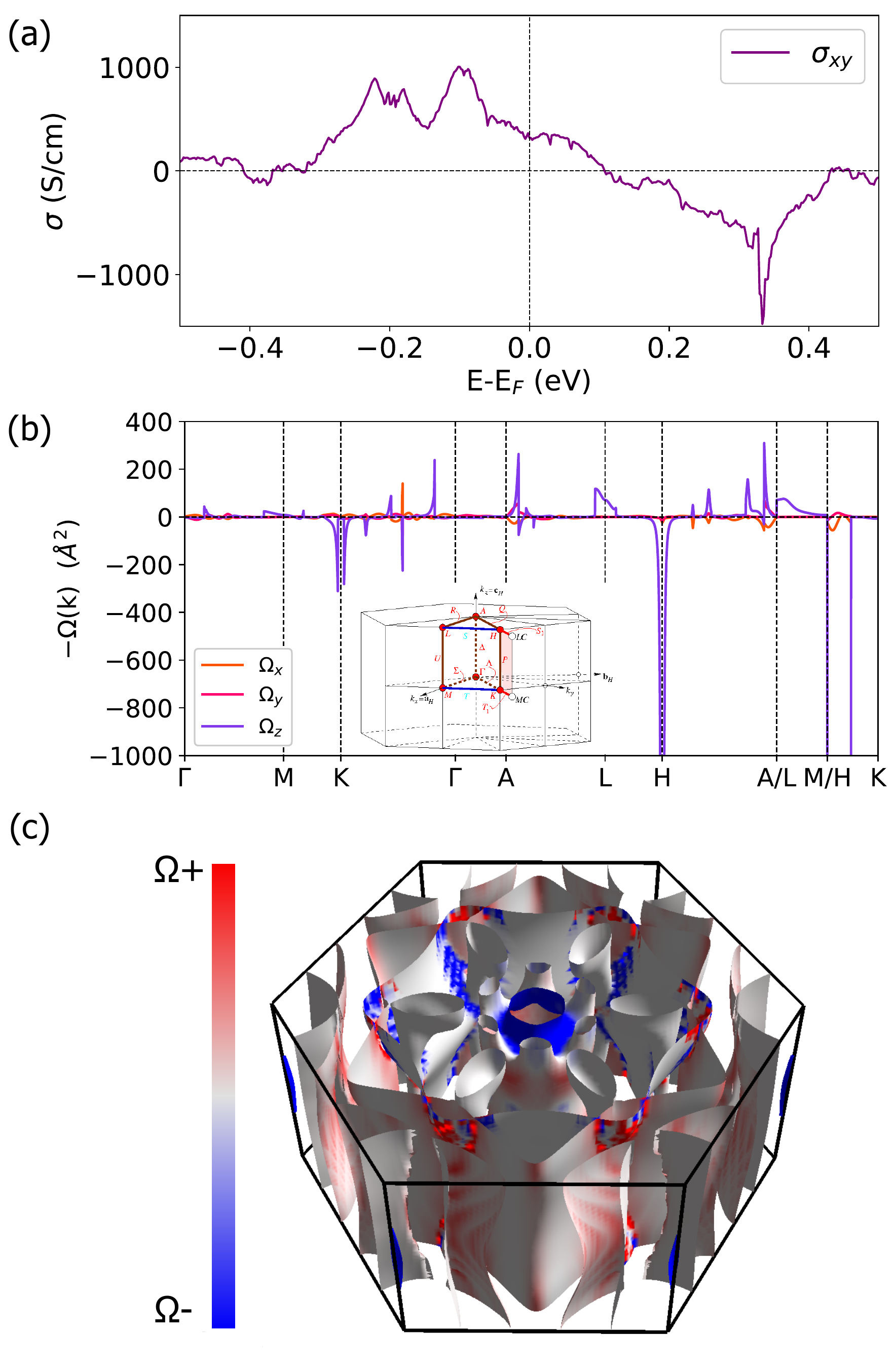}
 \caption{(Color online)  (a) Anomalous Hall conductivity, $\sigma_{xy}$ component, calculated (PBEsol) as a function of the chemical potential near the Fermi energy ($E_F$) for the ferromagnetic (001) KMn$_3$Sb$_5$ within the \emph{P6/mm'm'} magnetic space group (MSG.\,191.240). (b) Berry curvature, $\Omega_x(k)$, $\Omega_y(k)$, and $\Omega_z(k)$ components, calculated using DFT-PBEsol and integrated along the Brillouin zone $k$-path. In the inset, we display the hexagonal Brillouin zone for the SG.\,191 space group in which, the high-symmetry points and relevant $k$-paths are marked. (c) $\Omega_z${\bf (k)} component of Berry curvature projected onto the Fermi surface of ferromagnetic (001) KMn$_3$Sb$_5$ using a color map. Blue and red colors denote the negative and positive components of $\Omega_z${\bf (k)}, respectively.  
 }
 \label{F3}
\end{figure}

Fig.~\ref{F3}(a) shows the variation of $\sigma_{xy}$ as a function of the chemical potential near the Fermi level, calculated using the Kubo formula \cite{doi:10.1143/JPSJ.12.570,PhysRevB.64.014416}. See note~\footnote{the anomalous Hall conductivity component, $\sigma_{xy}$, has been computed by following the relationship: 
\begin{equation}\label{eq:ahc}
\sigma_{xy}=-\frac{e^2}{\hbar} \sum_n^{occ} \int_{BZ} \frac{d^3k}{(2\pi)^3} f_n({\bf{k}}) \Omega_{n,xy} (\bf{k}),
\end{equation}
where $\Omega_{xy}(\bf{k})$=$\sum_n^{occ} f_n(\bf{k})$$\Omega_{n,xy}(\bf{k})$ corresponds to the Berry curvature in the $xy$-plane and it is the result of the summation of all the occupied $n$-bands and $f_n\bf(k)$ represents the Fermi distribution. \textcolor{black}{In this calculations, the $\Omega_{n,xy}(\bf{k})$ Berry curvature is estimated such as:}
\textcolor{black}{
\begin{equation}
\Omega_{n,xy}(\textbf{k}) = -2i\hbar^2 \sum_{m \neq n} \frac{\left\langle\psi_{n,\textbf{k}}\left|v_x\right| \psi_{m,\textbf{k}}\right\rangle\left\langle\psi_{m,\textbf{k}}\left|v_y\right| \psi_{n,\textbf{k}}\right\rangle}{\left[E_m(\textbf{k})-E_n(\textbf{k})\right]^2}
\end{equation}}
In the latter equation, $\psi_{n,\textbf{k}}$ and $v_x$, $v_y$ are the Bloch functions and the velocity operators, respectively. In this step, an 320$\times$320$\times$320 $k$-mesh was used.}
for more details. 
As it can be observed, the AHC at the Fermi level is $\sigma_{xy}$ = 314 S$\cdot$cm$^{-1}$.
This is comparable to the values obtained in similar kagome compounds with large, or even giant, AHC such as 
$\sigma_{xy}$ = 380 S$\cdot$cm$^{-1}$ in LiMn$_6$Sn$_6$~\cite{PhysRevB.103.144410},  
$\sigma_{xy}$ = 1130 S$\cdot$cm$^{-1}$ in Co$_3$Sn$_2$S$_2$~\cite{Co3Sn2S2-kagome-NatPhys-2018}, 
and $\sigma_{xy}$ = $-$400 S$\cdot$cm$^{-1}$ in Fe$_3$Sn$_2$ \cite{PhysRevB.94.075135}.
Notably, by tuning the Fermi level in kagome metal KMn$_3$Sb$_5$, it is possible to achieve a $\sigma_{xy}$ value close to or even larger than 1000 S$\cdot$cm$^{-1}$.

In Fig.~\ref{F3}(b), we present the Berry curvature components $\Omega_{x}$, $\Omega_{y}$, and $\Omega_{z}$ calculated along the high-symmetry directions in the Brillouin zone. 
The largest contributions to the Berry curvature are obtained for $\Omega_{z}$ at the $H$-point and along the $H$--$K$ path. The values extend up to $-$7000 \r{A}$^2$, though in Fig.~\ref{F3}(b), the $y$-axis is displayed only up to $-$1000 \r{A}$^2$ to allow for the observation of smaller contributions along the other Brillouin zone paths. 
By correlating the Berry curvature with the electronic band structure calculated with SOC, we notice that there is a SOC-induced gapped node at the $H$ point near the Fermi energy (see Fig.~\ref{F2}), suggesting the presence of potential gapless Weyl nodes in the vicinity of the $H$ point. This explains the divergent Berry curvature at the Weyl nodes~\cite{RevModPhys.82.1959} which are the main source of the large AHC response in KMn$_3$Sb$_5$. 
Similar behavior has also been reported in antiperovskite compounds such as Mn$_3$NiN and V$_3$AuN~\cite{PhysRevB.106.195113, PhysRevMaterials.6.125003}.
Fig.~\ref{F3}(b) also displays the hexagonal Brillouin zone obtained for $P6/mmm$ space group in which, the high-symmetry points and relevant $k$-paths are marked. 
Fig.~\ref{F3}(c) shows the calculated Berry curvature $\Omega_z$({\bf k}) projected onto the Fermi surface of KMn$_3$Sb$_5$. Here, the red and blue colors denote the positive and negative $\Omega_z$({\bf k}) values. Throughout the entire Brillouin zone, multiple strong concentrations of Berry curvature can be observed.

%%%%%%%%%%%%%%%%
%%%%%%%%%%%%%%%%

%%%%%%%%%%%%%%%%%%%%%%%%%%%%%%%%%%%%%%%%%%%%%%%%
\subsection{Role of Electronic Correlations:}

As shown in previous reports~\cite{PhysRevB.105.235145, PhysRevResearch.5.L012008, PhysRevLett.127.177001, FLOREZGOMEZ2022169813}, the kagome lattices with magnetic cations, especially cations having partially-filled $3d$ valence shell, exhibit strong electronic-correlation phenomenon that leads to the emergence of exciting electronic properties. 
Therefore, to investigate the effects of the onsite Coulomb interaction on Mn-$3d$ electrons and its impact on the structural, phononic, and electronic properties, we employed DFT+$U$ calculations which capture the electronic-correlations effects at the mean-field level~\cite{Liechtenstein1995}. 
This analysis is particularly important due to the lack of experimental reports on KMn$_3$Sb$_5$, and it may help experimentalists in identifying suitable observables for future investigations.

In Table \ref{tab:1}, we present the evolution of the lattice parameters, the magnetic moment per Mn atom, and anomalous Hall conductivity ($\sigma_{xy}$) values as a function of the $+U$ Coulomb correction parameter applied to Mn-$3d$ electrons. 
Our results reveal an increase in the in-plane lattice parameters ($a=b$) while decrease in the out-of-plane ($c$) lattice parameter with increasing $U$. 
%As it can be observed, the increase in the Coulomb repulsion value, considerably increases the in-plane $a$ lattice parameter, from 5.337 \r{A} to 5.602 \r{A} when going from $U$ = 0.0 eV to $U$ = 4.0 eV. 
This demonstrates the dominant role of the Mn--Mn bondings and interactions within the kagome lattice. 
The observed expansion in the in-plane lattice parameters is in agreement with the previous observation of the strong spin-lattice coupling related to the negative thermal expansion in the kagome antiperovskite Mn$_3$NiN \cite{FLOREZGOMEZ2022169813}. It is worth noting that with increasing $U$ value, some phonon modes (especially the in-plane vibrational modes) soften due to the increased in-plane lattice parameters. However, no unstable phonon modes are observed within the range of considered $U$ values. 
%Consequently, the $c$ lattice parameter is reduced from 9.028 \r{A} to 8.606 \r{A}. 
%
As expected, the magnetic moment per Mn-site is increased due to the increased electron localization governed by the Coulomb term. 
Interestingly, the $\sigma_{xy}$ component remains substantial at all the considered $U$ values.
%with a minimum value of 217 S$\cdot$cm$^{-1}$ at $U$ = 1.0 eV.} 

Finally, we fully relax the KMn$_3$Sb$_5$ FM (001) structure using the meta-GGA \textsc{scan} \cite{PhysRevLett.115.036402} and r$^{2}$\textsc{scan} \cite{doi:10.1021/acs.jpclett.0c02405} functionals,  aiming to obtain more precise values of the interesting observables.  
We find that the magnetic moment is 3.022 $\mu_B$$\cdot$Mn$^{-1}$ and 3.229 $\mu_B$$\cdot$Mn$^{-1}$ for the  \textsc{scan} and  r$^{2}$\textsc{scan} functionals, respectively.
Moreover, the lattice parameters are $a$ = 5.435 \r{A} and $c$ = 9.153 \r{A} for the \textsc{scan} functional whereas $a$ = 5.512 \r{A} and $c$ = 9.042 \r{A} for the r$^{2}$\textsc{scan} functional. 
Using these meta-GGA values as observables, we suggest that a $U$ value close to 2\,eV could reasonably reproduce the structural and electronic properties of KMn$_3$Sb$_5$. 
However, it is important to note that more systematic studies are needed to fully comprehend the electron-correlation effects in this studied kagome metal.
%
%for lattice parameters, we estimate $U$ for Mn-$3d$ electrons in FM (001) KMn$_3$Sb$_5$ to be in $2-3$\,eV range. 
%We note that the magnetic moment in the \textsc{scan}  and  \textsc{r2scan} functionals are often overestimated~\cite{PhysRevB.98.094413,PhysRevLett.121.207201,PhysRevB.100.045126,doi:10.1021/acsmaterialsau.2c00059}. 
%However, we suggest that a $U$ value close to 1\,eV could fairly reproduce the structure and electronic properties in KMn$_3$Sb$_5$ kgome metal. 
%Nevertheless, more systematic studies are needed in order to fully comprehend the electron-correlation effects in the studied kagome metal. 

%%%%%%%%%%%%%%%%
\begin{table}[t]
\centering
\caption{Lattice parameters ($a=b$, and $c$), magnetic moment, and $\sigma_{xy}$ AHC component as a function of different $U$ values in our PBEsol+$U$ calculations. Values obtained using the meta-GGA \textsc{scan}, and $r^{2}$\textsc{scan} functionals are also listed. 
All the structures were fully relaxed considering the ground state (001) noncollinear ferromagnetic ordering. }
\begin{tabular}{c | c  c  c  c  c  c}
\hline
\hline
XC-type &  $a$ (\r{A}) &  $c$ (\r{A})  &  $m$ ($\mu_B$$\cdot$Mn$^{-1}$) & $\sigma_{xy}$ (S$\cdot$cm$^{-1}$)  \rule[-1ex]{0pt}{3.5ex} \\
\hline
PS+$U$=0.0 eV &  5.337 &  9.028 & 2.458  & 314 \rule[-1ex]{0pt}{3.5ex}  \\
PS+$U$=1.0 eV &  5.372 &  9.014 & 2.758  & 217 \rule[-1ex]{0pt}{3.5ex}  \\
PS+$U$=2.0 eV &  5.470 &  8.875 & 3.312  & 241 \rule[-1ex]{0pt}{3.5ex}  \\
PS+$U$=3.0 eV &  5.602 &  8.606 & 3.743  & 172 \rule[-1ex]{0pt}{3.5ex}  \\
%PS+$U$=4.0 eV &  5.720 &  8.510 & 4.184  & ??? \rule[-1ex]{0pt}{3.5ex}  \\
\hline
\textsc{scan} &  5.435 &  9.153 & 3.022 & ---  \rule[-1ex]{0pt}{3.5ex}  \\
r$^{2}$\textsc{scan} &  5.512 &  9.042 & 3.229 & ---  \rule[-1ex]{0pt}{3.5ex}  \\
\hline
\hline
\end{tabular}
\label{tab:1}
\end{table}
%%%%%%%%%%%%%%%%

%After considering an A-type antiferromagnetic ordering, $i.e$ ferromagnetic Kagome in-layer coupled antiferromagnetically with the subsequent layer in the $z$-axis, we found that $\Delta$$E$ = $E_{A-AFM}$ $-$ $E_{FM}$ = $-$3.75 meV. The latter is similar to the magnetic coupling observed, experimentally and theoretically, in the Kagome FeSn compound \cite{JBernhard_1984}.

%%%%%%%%%%%%%%%%%%%%%%%%%%%%%%%%%%%%%%%%%%%%%%%%
\section{Conclusions} \label{secIV}

In this study, we performed comprehensive first-principles DFT calculations to investigate the properties of the novel Mn-based kagome metal KMn$_3$Sb$_5$. 
Our results indicate that KMn$_3$Sb$_5$ is both structurally and vibrationally stable, with no observed unstable phonon modes.  
This is in contrast to the KV$_3$Sb$_5$ compound in which, phonon instability leads to the appearance of CDW phase. 
A detailed investigation of the possible candidate magnetic configurations (noncollinear FM as well as AFM) reveal the (001) ferromagnetic ordering as the ground state in KMn$_3$Sb$_5$, although various other magnetically-frustrated kagome metals favor noncollinear chiral antiferromagnetic orderings $\Gamma_{4g}$ and $\Gamma_{5g}$, similar to the case of Mn$_3$NiN antiperovskite, due to the inherent magnetic frustration. 
The electronic structure shows multiple nodal crossings associated with Weyl nodes near the Fermi energy, indicating topologically nontrivial features present in the studied material. 
Notably, the ferromagnetic order breaks the time-reversal symmetry, resulting in a tangible anomalous Hall conductivity response. We find a value of $\sigma_{xy}$ = 314 S$\cdot$cm$^{-1}$, which moderately varies with increasing onsite Coloumb parameter $U$ in our PBEsol+$U$ calculations. 
We find that the predicted $\sigma_{xy}$ in KMn$_3$Sb$_5$ to be considerably large when compared to other similar kagome compounds. 
We suggest a $U$ value in the range of 2-3\,eV to reasonably reproduce the properties of the ferromagnetic kagome metal KMn$_3$Sb$_5$.
Our findings are expected to motivate experimentalists to pursue the synthesis, realization, and subsequent confirmation of the predicted properties in KMn$_3$Sb$_5$.

%%%%%%%%%%%%%%%%%%%%%%%%%%%%%%%%%%%%%%%%%%%%%%%%
\section*{Acknowledgments}
Calculations presented in this article were carried out using the GridUIS-2 experimental testbed and Center for Integrated Research Computing (CIRC) facilities at the University of Rochester. 
The GridUIS-2 testbed was developed under the Universidad Industrial de Santander (SC3-UIS) High Performance and Scientific Computing Centre with support from UIS Vicerrector\'ia de Investigaci\'on y Extensi\'on (VIE-UIS) and several UIS research groups.
We also acknowledge the computational resources awarded by XSEDE, a project supported by National Science Foundation grant number ACI-1053575. The authors also acknowledge the support from the Texas Advances Computer Center (with the Stampede2 and Bridges supercomputers).

\appendix

%%%%%%%%%%%%%%%%%%%%
\section{Phonons under the $+U$ correction.}
\label{secA}

%%%%%%%%%%%%%%%%
\begin{figure}[!h]
 \centering
 \includegraphics[width=8.8cm,keepaspectratio=true]{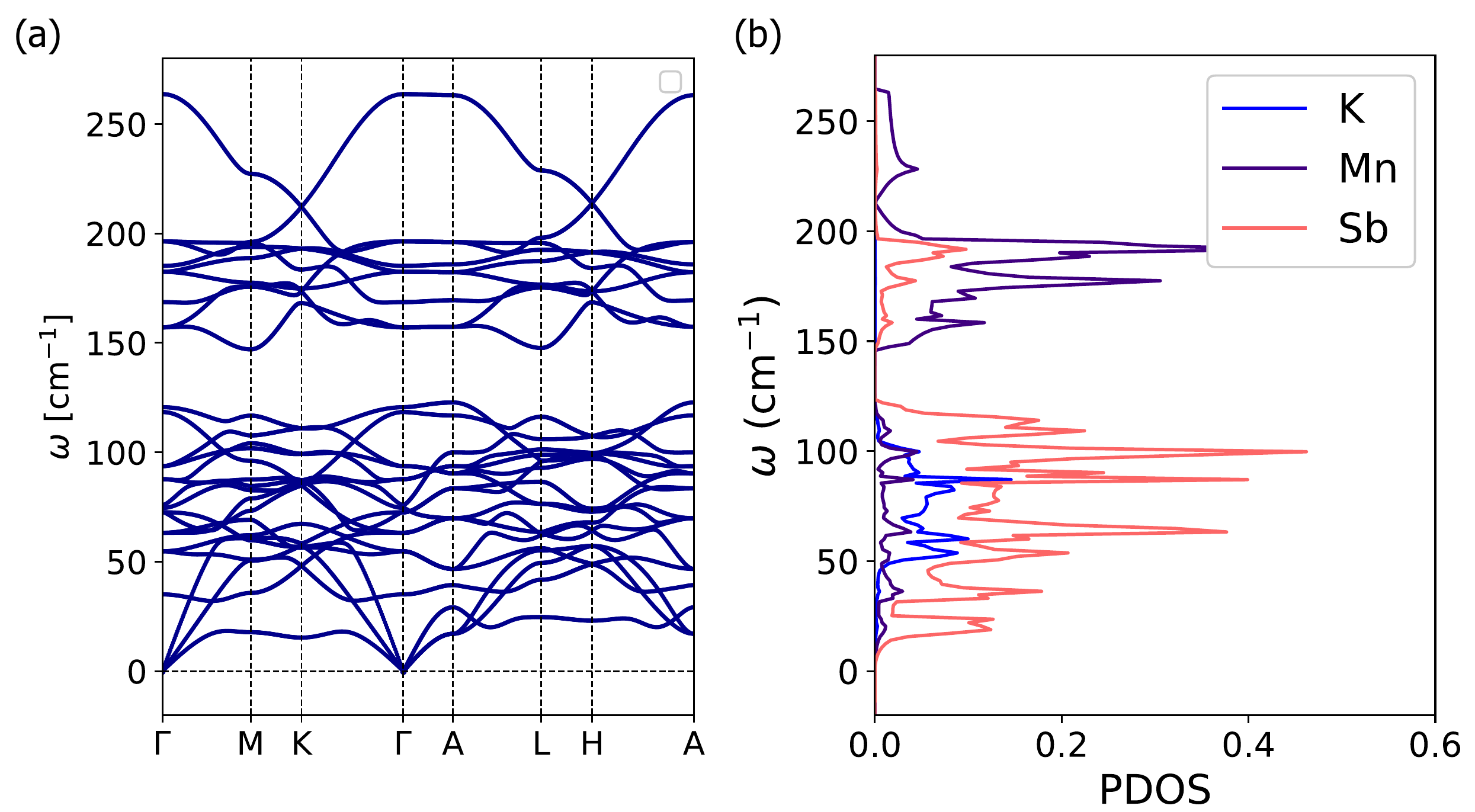}
 \caption{(Color online)  (a) Full phonon-dispersion curve and (b) Atom-projected phonon density of states obtained at the Coulomb $+U$ value of 3.0 eV as a correction in the 3$d$:Mn states at the KMn$_3$Sb$_5$ Kagome compound.}
 \label{F2a}
\end{figure} 
%%%%%%%%%%%%%%%%

%%%%%%%%%%%%%%%%
\begin{figure}[!h]
 \centering
 \includegraphics[width=8.6cm,keepaspectratio=true]{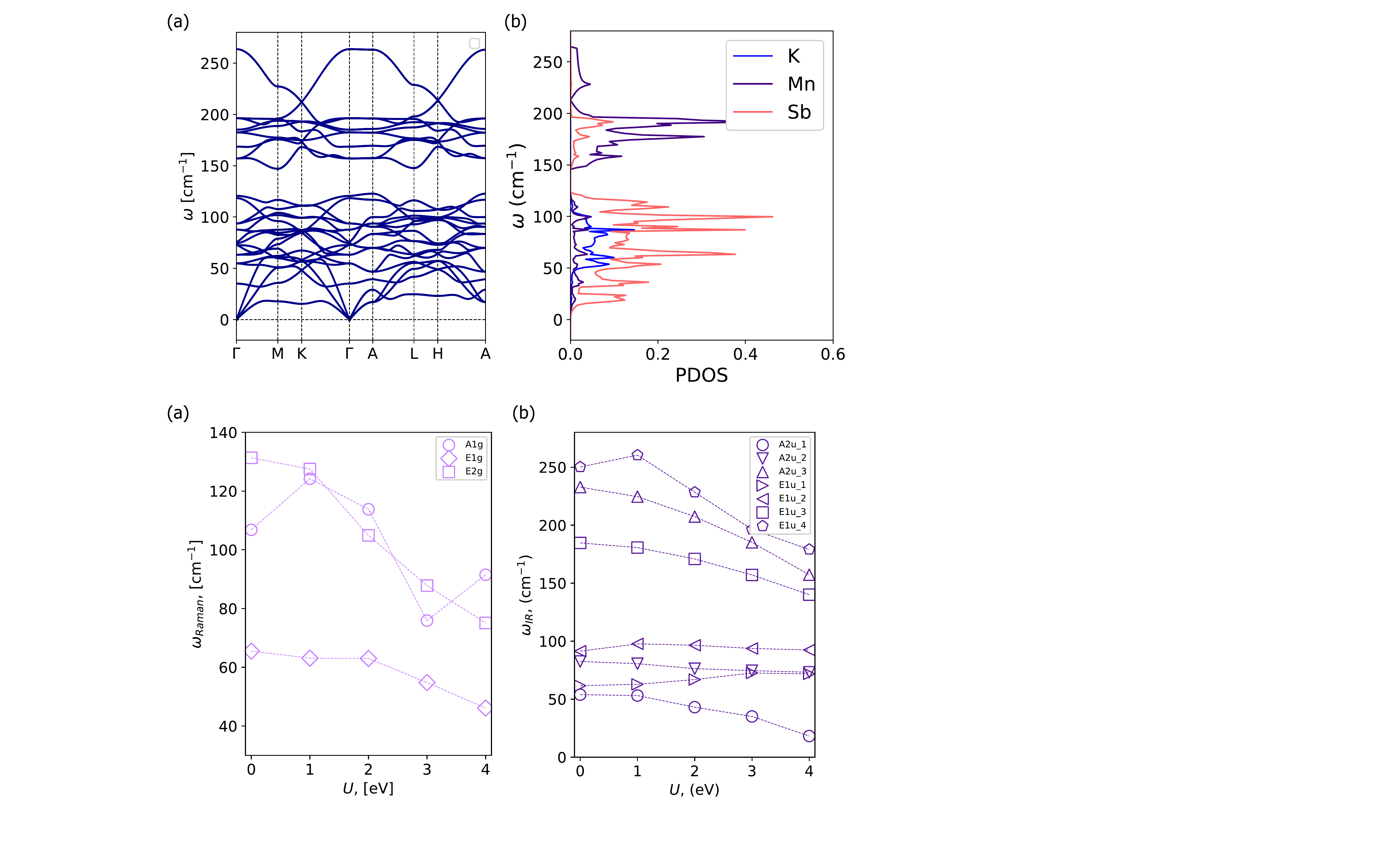}
 \caption{(Color online) BZ zone center Raman and IR active modes as a function of the $+U$ parameter in the KMn$_3$Sb$_5$.}
 \label{F2a-2}
\end{figure} 
%%%%%%%%%%%%%%%%

%%%%%%%%%%%%%%%%%%%%
\section{Electronic bands under the $+U$ correction.}
\label{secB}

%%%%%%%%%%%%%%%%
\begin{figure}[!h]
 \centering
 \includegraphics[width=8.6cm,keepaspectratio=true]{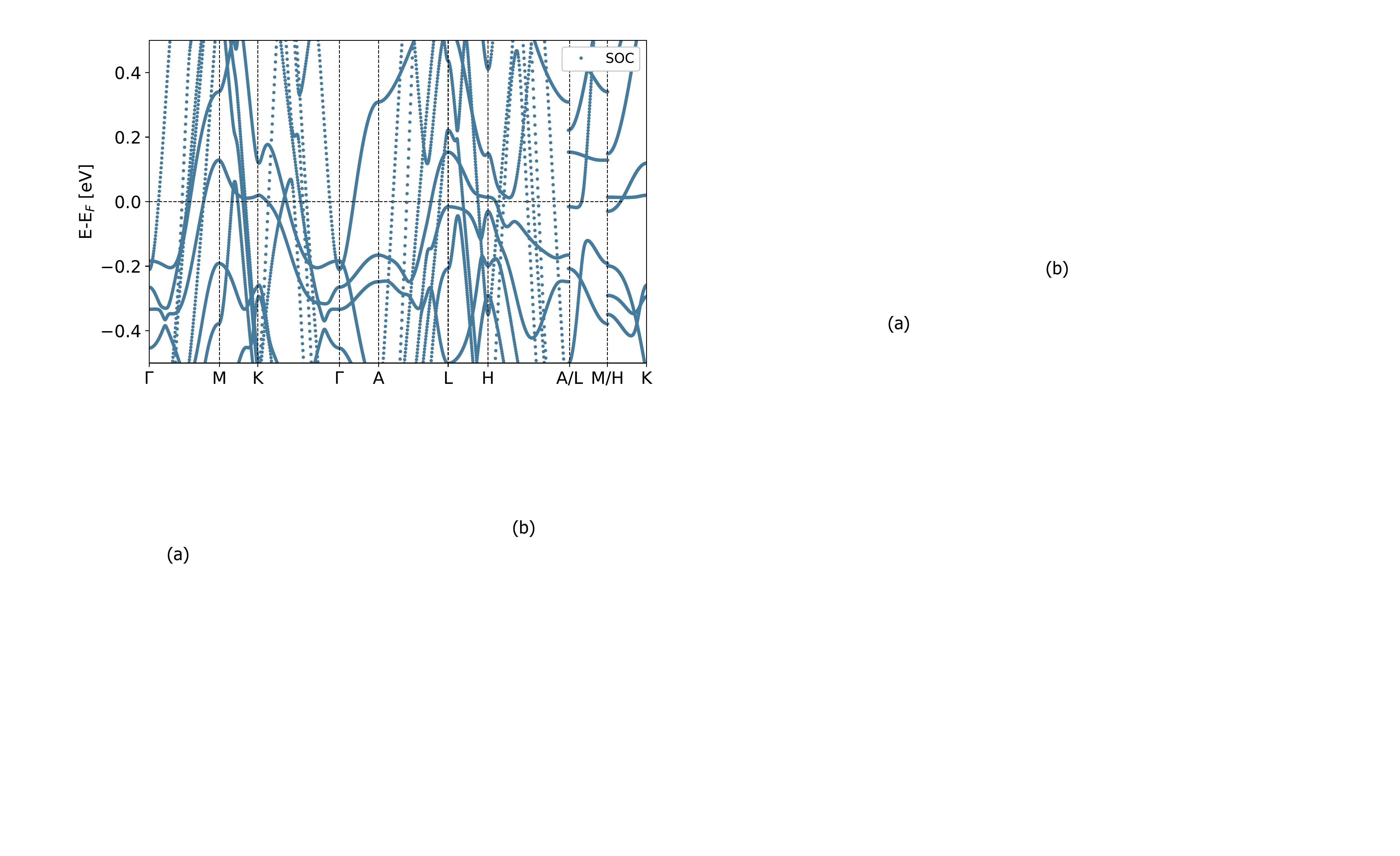}
 \caption{(Color online) (a) Full dispersion bands obtained at the Coulomb $+U$ value of 3.0 eV as a correction in the 3$d$:Mn states at the KMn$_3$Sb$_5$ Kagome compound.}
 \label{F3a}
\end{figure} 
%%%%%%%%%%%%%%%%

%%%%%%%%%%%%%%%
\section{Spin-polarized band structure.}
\label{secC}

%%%%%%%%%%%%%%%%
\begin{figure}[!h]
 \centering
 \includegraphics[width=8.6cm,keepaspectratio=true]{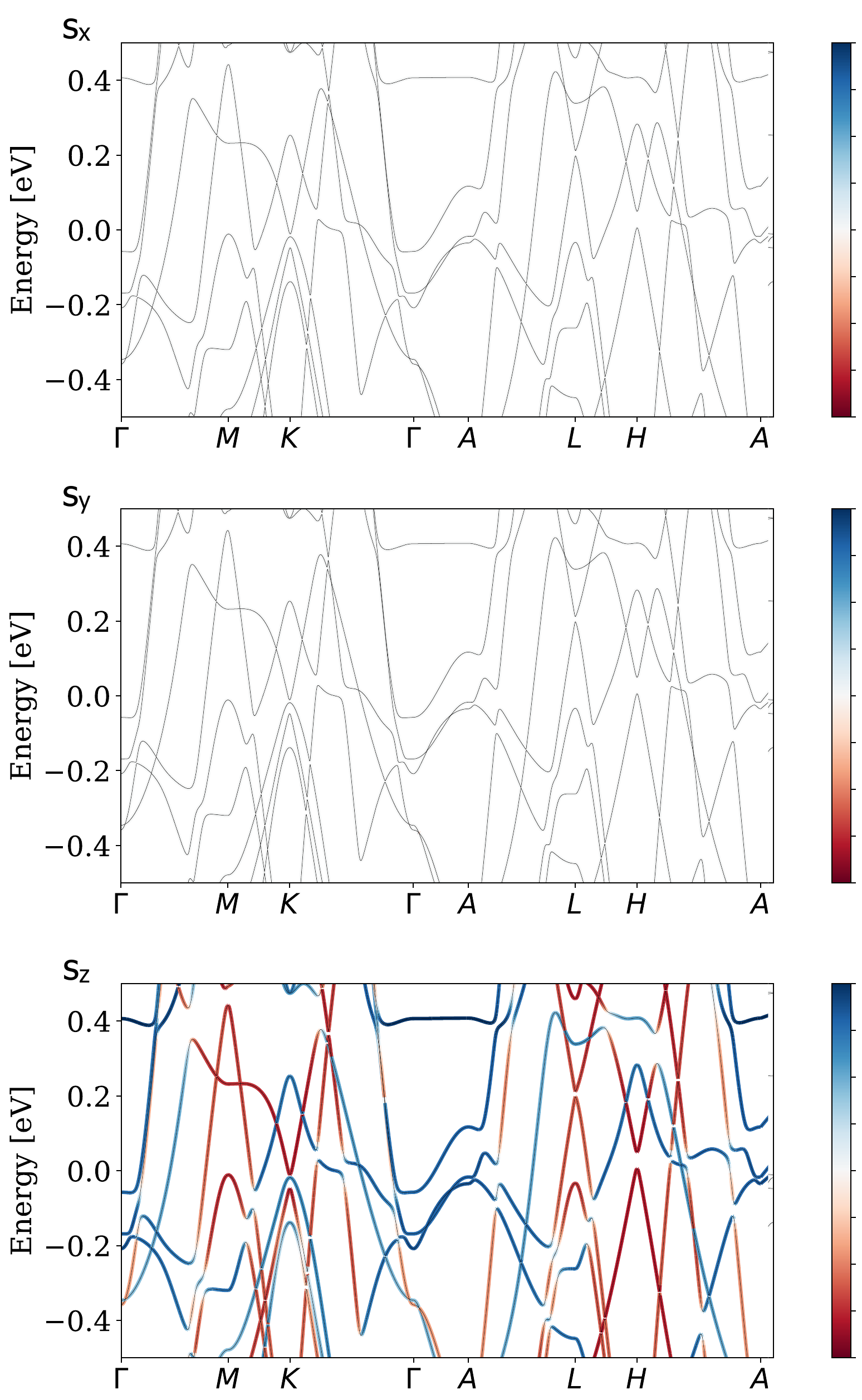}
 \caption{(Color online)  Spin-polarized electronic band structure showing the $S_x$, $S_y$, and $S_z$ spin components, respectively. Here, it can be noted that the contributions from the $S_x$ and $S_y$ spin projections are negligible, and only, the $S_z$ component is appreciated in the E$_F$ proximity. Additionally, the Weyl nodes can be identified in the path between the M--K and K--$\Gamma$ points and consequently, in the associated point, L--H and H--A in the (0,\,0,\,1/2) plane. Here, the blue and red colors denote the up- and down-spin projections, respectively. 
 }
 \label{F1a}
\end{figure} 
%%%%%%%%%%%%%%%%

\bibliography{library}

\begin{thebibliography}{63}%
\makeatletter
\providecommand \@ifxundefined [1]{%
 \@ifx{#1\undefined}
}%
\providecommand \@ifnum [1]{%
 \ifnum #1\expandafter \@firstoftwo
 \else \expandafter \@secondoftwo
 \fi
}%
\providecommand \@ifx [1]{%
 \ifx #1\expandafter \@firstoftwo
 \else \expandafter \@secondoftwo
 \fi
}%
\providecommand \natexlab [1]{#1}%
\providecommand \enquote  [1]{``#1''}%
\providecommand \bibnamefont  [1]{#1}%
\providecommand \bibfnamefont [1]{#1}%
\providecommand \citenamefont [1]{#1}%
\providecommand \href@noop [0]{\@secondoftwo}%
\providecommand \href [0]{\begingroup \@sanitize@url \@href}%
\providecommand \@href[1]{\@@startlink{#1}\@@href}%
\providecommand \@@href[1]{\endgroup#1\@@endlink}%
\providecommand \@sanitize@url [0]{\catcode `\\12\catcode `\$12\catcode
  `\&12\catcode `\#12\catcode `\^12\catcode `\_12\catcode `\%12\relax}%
\providecommand \@@startlink[1]{}%
\providecommand \@@endlink[0]{}%
\providecommand \url  [0]{\begingroup\@sanitize@url \@url }%
\providecommand \@url [1]{\endgroup\@href {#1}{\urlprefix }}%
\providecommand \urlprefix  [0]{URL }%
\providecommand \Eprint [0]{\href }%
\providecommand \doibase [0]{https://doi.org/}%
\providecommand \selectlanguage [0]{\@gobble}%
\providecommand \bibinfo  [0]{\@secondoftwo}%
\providecommand \bibfield  [0]{\@secondoftwo}%
\providecommand \translation [1]{[#1]}%
\providecommand \BibitemOpen [0]{}%
\providecommand \bibitemStop [0]{}%
\providecommand \bibitemNoStop [0]{.\EOS\space}%
\providecommand \EOS [0]{\spacefactor3000\relax}%
\providecommand \BibitemShut  [1]{\csname bibitem#1\endcsname}%
\let\auto@bib@innerbib\@empty
%</preamble>
\bibitem [{\citenamefont {Mekata}(2003)}]{doi:10.1063/1.1564329}%
  \BibitemOpen
  \bibfield  {author} {\bibinfo {author} {\bibfnamefont {M.}~\bibnamefont
  {Mekata}},\ }\bibfield  {title} {\bibinfo {title} {Kagome: The story of the
  basketweave lattice},\ }\href {https://doi.org/10.1063/1.1564329} {\bibfield
  {journal} {\bibinfo  {journal} {Physics Today}\ }\textbf {\bibinfo {volume}
  {56}},\ \bibinfo {pages} {12} (\bibinfo {year} {2003})},\ \Eprint
  {https://arxiv.org/abs/https://doi.org/10.1063/1.1564329}
  {https://doi.org/10.1063/1.1564329} \BibitemShut {NoStop}%
\bibitem [{\citenamefont {Kang}\ \emph {et~al.}(2020)\citenamefont {Kang},
  \citenamefont {Ye}, \citenamefont {Fang}, \citenamefont {You}, \citenamefont
  {Levitan}, \citenamefont {Han}, \citenamefont {Facio}, \citenamefont
  {Jozwiak}, \citenamefont {Bostwick}, \citenamefont {Rotenberg}, \citenamefont
  {Chan}, \citenamefont {McDonald}, \citenamefont {Graf}, \citenamefont
  {Kaznatcheev}, \citenamefont {Vescovo}, \citenamefont {Bell}, \citenamefont
  {Kaxiras}, \citenamefont {van~den Brink}, \citenamefont {Richter},
  \citenamefont {Prasad~Ghimire}, \citenamefont {Checkelsky},\ and\
  \citenamefont {Comin}}]{FeSn-NatMat-2020}%
  \BibitemOpen
  \bibfield  {author} {\bibinfo {author} {\bibfnamefont {M.}~\bibnamefont
  {Kang}}, \bibinfo {author} {\bibfnamefont {L.}~\bibnamefont {Ye}}, \bibinfo
  {author} {\bibfnamefont {S.}~\bibnamefont {Fang}}, \bibinfo {author}
  {\bibfnamefont {J.-S.}\ \bibnamefont {You}}, \bibinfo {author} {\bibfnamefont
  {A.}~\bibnamefont {Levitan}}, \bibinfo {author} {\bibfnamefont
  {M.}~\bibnamefont {Han}}, \bibinfo {author} {\bibfnamefont {J.~I.}\
  \bibnamefont {Facio}}, \bibinfo {author} {\bibfnamefont {C.}~\bibnamefont
  {Jozwiak}}, \bibinfo {author} {\bibfnamefont {A.}~\bibnamefont {Bostwick}},
  \bibinfo {author} {\bibfnamefont {E.}~\bibnamefont {Rotenberg}}, \bibinfo
  {author} {\bibfnamefont {M.~K.}\ \bibnamefont {Chan}}, \bibinfo {author}
  {\bibfnamefont {R.~D.}\ \bibnamefont {McDonald}}, \bibinfo {author}
  {\bibfnamefont {D.}~\bibnamefont {Graf}}, \bibinfo {author} {\bibfnamefont
  {K.}~\bibnamefont {Kaznatcheev}}, \bibinfo {author} {\bibfnamefont
  {E.}~\bibnamefont {Vescovo}}, \bibinfo {author} {\bibfnamefont {D.~C.}\
  \bibnamefont {Bell}}, \bibinfo {author} {\bibfnamefont {E.}~\bibnamefont
  {Kaxiras}}, \bibinfo {author} {\bibfnamefont {J.}~\bibnamefont {van~den
  Brink}}, \bibinfo {author} {\bibfnamefont {M.}~\bibnamefont {Richter}},
  \bibinfo {author} {\bibfnamefont {M.}~\bibnamefont {Prasad~Ghimire}},
  \bibinfo {author} {\bibfnamefont {J.~G.}\ \bibnamefont {Checkelsky}},\ and\
  \bibinfo {author} {\bibfnamefont {R.}~\bibnamefont {Comin}},\ }\bibfield
  {title} {\bibinfo {title} {Dirac fermions and flat bands in the ideal kagome
  metal fesn},\ }\href {https://doi.org/10.1038/s41563-019-0531-0} {\bibfield
  {journal} {\bibinfo  {journal} {Nature Materials}\ }\textbf {\bibinfo
  {volume} {19}},\ \bibinfo {pages} {163} (\bibinfo {year} {2020})}\BibitemShut
  {NoStop}%
\bibitem [{\citenamefont {Liu}\ \emph {et~al.}(2018{\natexlab{a}})\citenamefont
  {Liu}, \citenamefont {Sun}, \citenamefont {Kumar}, \citenamefont {Muechler},
  \citenamefont {Sun}, \citenamefont {Jiao}, \citenamefont {Yang},
  \citenamefont {Liu}, \citenamefont {Liang}, \citenamefont {Xu}, \citenamefont
  {Kroder}, \citenamefont {S{\"u}{\ss}}, \citenamefont {Borrmann},
  \citenamefont {Shekhar}, \citenamefont {Wang}, \citenamefont {Xi},
  \citenamefont {Wang}, \citenamefont {Schnelle}, \citenamefont {Wirth},
  \citenamefont {Chen}, \citenamefont {Goennenwein},\ and\ \citenamefont
  {Felser}}]{Co3Sn2S2-kagome-NatPhys-2018}%
  \BibitemOpen
  \bibfield  {author} {\bibinfo {author} {\bibfnamefont {E.}~\bibnamefont
  {Liu}}, \bibinfo {author} {\bibfnamefont {Y.}~\bibnamefont {Sun}}, \bibinfo
  {author} {\bibfnamefont {N.}~\bibnamefont {Kumar}}, \bibinfo {author}
  {\bibfnamefont {L.}~\bibnamefont {Muechler}}, \bibinfo {author}
  {\bibfnamefont {A.}~\bibnamefont {Sun}}, \bibinfo {author} {\bibfnamefont
  {L.}~\bibnamefont {Jiao}}, \bibinfo {author} {\bibfnamefont {S.-Y.}\
  \bibnamefont {Yang}}, \bibinfo {author} {\bibfnamefont {D.}~\bibnamefont
  {Liu}}, \bibinfo {author} {\bibfnamefont {A.}~\bibnamefont {Liang}}, \bibinfo
  {author} {\bibfnamefont {Q.}~\bibnamefont {Xu}}, \bibinfo {author}
  {\bibfnamefont {J.}~\bibnamefont {Kroder}}, \bibinfo {author} {\bibfnamefont
  {V.}~\bibnamefont {S{\"u}{\ss}}}, \bibinfo {author} {\bibfnamefont
  {H.}~\bibnamefont {Borrmann}}, \bibinfo {author} {\bibfnamefont
  {C.}~\bibnamefont {Shekhar}}, \bibinfo {author} {\bibfnamefont
  {Z.}~\bibnamefont {Wang}}, \bibinfo {author} {\bibfnamefont {C.}~\bibnamefont
  {Xi}}, \bibinfo {author} {\bibfnamefont {W.}~\bibnamefont {Wang}}, \bibinfo
  {author} {\bibfnamefont {W.}~\bibnamefont {Schnelle}}, \bibinfo {author}
  {\bibfnamefont {S.}~\bibnamefont {Wirth}}, \bibinfo {author} {\bibfnamefont
  {Y.}~\bibnamefont {Chen}}, \bibinfo {author} {\bibfnamefont {S.~T.~B.}\
  \bibnamefont {Goennenwein}},\ and\ \bibinfo {author} {\bibfnamefont
  {C.}~\bibnamefont {Felser}},\ }\bibfield  {title} {\bibinfo {title} {Giant
  anomalous hall effect in a ferromagnetic kagome-lattice semimetal},\ }\href
  {https://doi.org/10.1038/s41567-018-0234-5} {\bibfield  {journal} {\bibinfo
  {journal} {Nature Physics}\ }\textbf {\bibinfo {volume} {14}},\ \bibinfo
  {pages} {1125} (\bibinfo {year} {2018}{\natexlab{a}})}\BibitemShut {NoStop}%
\bibitem [{\citenamefont {Arachchige}\ \emph {et~al.}(2022)\citenamefont
  {Arachchige}, \citenamefont {Meier}, \citenamefont {Marshall}, \citenamefont
  {Matsuoka}, \citenamefont {Xue}, \citenamefont {McGuire}, \citenamefont
  {Hermann}, \citenamefont {Cao},\ and\ \citenamefont
  {Mandrus}}]{PhysRevLett.129.216402}%
  \BibitemOpen
  \bibfield  {author} {\bibinfo {author} {\bibfnamefont {H.~W.~S.}\
  \bibnamefont {Arachchige}}, \bibinfo {author} {\bibfnamefont {W.~R.}\
  \bibnamefont {Meier}}, \bibinfo {author} {\bibfnamefont {M.}~\bibnamefont
  {Marshall}}, \bibinfo {author} {\bibfnamefont {T.}~\bibnamefont {Matsuoka}},
  \bibinfo {author} {\bibfnamefont {R.}~\bibnamefont {Xue}}, \bibinfo {author}
  {\bibfnamefont {M.~A.}\ \bibnamefont {McGuire}}, \bibinfo {author}
  {\bibfnamefont {R.~P.}\ \bibnamefont {Hermann}}, \bibinfo {author}
  {\bibfnamefont {H.}~\bibnamefont {Cao}},\ and\ \bibinfo {author}
  {\bibfnamefont {D.}~\bibnamefont {Mandrus}},\ }\bibfield  {title} {\bibinfo
  {title} {Charge density wave in kagome lattice intermetallic
  ${\mathrm{scv}}_{6}{\mathrm{sn}}_{6}$},\ }\href
  {https://doi.org/10.1103/PhysRevLett.129.216402} {\bibfield  {journal}
  {\bibinfo  {journal} {Phys. Rev. Lett.}\ }\textbf {\bibinfo {volume} {129}},\
  \bibinfo {pages} {216402} (\bibinfo {year} {2022})}\BibitemShut {NoStop}%
\bibitem [{\citenamefont {Wang}\ \emph {et~al.}(2019)\citenamefont {Wang},
  \citenamefont {Zhang}, \citenamefont {Zhu}, \citenamefont {L{\"{u}}},
  \citenamefont {Li}, \citenamefont {Zou},\ and\ \citenamefont
  {Zhao}}]{Wang2019}%
  \BibitemOpen
  \bibfield  {author} {\bibinfo {author} {\bibfnamefont {Y.}~\bibnamefont
  {Wang}}, \bibinfo {author} {\bibfnamefont {H.}~\bibnamefont {Zhang}},
  \bibinfo {author} {\bibfnamefont {J.}~\bibnamefont {Zhu}}, \bibinfo {author}
  {\bibfnamefont {X.}~\bibnamefont {L{\"{u}}}}, \bibinfo {author}
  {\bibfnamefont {S.}~\bibnamefont {Li}}, \bibinfo {author} {\bibfnamefont
  {R.}~\bibnamefont {Zou}},\ and\ \bibinfo {author} {\bibfnamefont
  {Y.}~\bibnamefont {Zhao}},\ }\bibfield  {title} {\bibinfo {title}
  {{Antiperovskites with Exceptional Functionalities}},\ }\href
  {https://doi.org/10.1002/adma.201905007} {\bibfield  {journal} {\bibinfo
  {journal} {Advanced Materials}\ }\textbf {\bibinfo {volume} {1905007}},\
  \bibinfo {pages} {1} (\bibinfo {year} {2019})}\BibitemShut {NoStop}%
\bibitem [{\citenamefont {Garcia-Castro}\ \emph {et~al.}(2020)\citenamefont
  {Garcia-Castro}, \citenamefont {Ospina},\ and\ \citenamefont
  {Quintero}}]{Garcia-Castro2020}%
  \BibitemOpen
  \bibfield  {author} {\bibinfo {author} {\bibfnamefont {A.~C.}\ \bibnamefont
  {Garcia-Castro}}, \bibinfo {author} {\bibfnamefont {R.}~\bibnamefont
  {Ospina}},\ and\ \bibinfo {author} {\bibfnamefont {J.~H.}\ \bibnamefont
  {Quintero}},\ }\bibfield  {title} {\bibinfo {title} {{Octahedral distortion
  and electronic properties of the antiperovskite oxide {Ba$_3$SiO}: First
  principles study}},\ }\href {https://doi.org/10.1016/j.jpcs.2019.109126}
  {\bibfield  {journal} {\bibinfo  {journal} {Journal of Physics and Chemistry
  of Solids}\ }\textbf {\bibinfo {volume} {136}},\ \bibinfo {pages} {109126}
  (\bibinfo {year} {2020})}\BibitemShut {NoStop}%
\bibitem [{\citenamefont {Garcia-Castro}\ \emph {et~al.}(2019)\citenamefont
  {Garcia-Castro}, \citenamefont {{Quintero Orozco}},\ and\ \citenamefont
  {{Paez Gonzalez}}}]{Garcia-Castro2019}%
  \BibitemOpen
  \bibfield  {author} {\bibinfo {author} {\bibfnamefont {A.~C.}\ \bibnamefont
  {Garcia-Castro}}, \bibinfo {author} {\bibfnamefont {J.~H.}\ \bibnamefont
  {{Quintero Orozco}}},\ and\ \bibinfo {author} {\bibfnamefont {C.~J.}\
  \bibnamefont {{Paez Gonzalez}}},\ }\bibfield  {title} {\bibinfo {title}
  {{Hybrid-improper ferroelectric behavior in {Ba$_3$SiO}/{Ba$_3$GeO} oxide
  antiperovskite superlattices}},\ }\href
  {https://doi.org/10.1140/epjb/e2019-100175-1} {\bibfield  {journal} {\bibinfo
   {journal} {European Physical Journal B}\ }\textbf {\bibinfo {volume} {92}},\
  \bibinfo {pages} {2} (\bibinfo {year} {2019})}\BibitemShut {NoStop}%
\bibitem [{\citenamefont {Torres-Amaris}\ \emph {et~al.}(2022)\citenamefont
  {Torres-Amaris}, \citenamefont {Bautista-Hernandez}, \citenamefont
  {Gonz\'alez-Hern\'andez}, \citenamefont {Romero},\ and\ \citenamefont
  {Garcia-Castro}}]{PhysRevB.106.195113}%
  \BibitemOpen
  \bibfield  {author} {\bibinfo {author} {\bibfnamefont {D.}~\bibnamefont
  {Torres-Amaris}}, \bibinfo {author} {\bibfnamefont {A.}~\bibnamefont
  {Bautista-Hernandez}}, \bibinfo {author} {\bibfnamefont {R.}~\bibnamefont
  {Gonz\'alez-Hern\'andez}}, \bibinfo {author} {\bibfnamefont {A.~H.}\
  \bibnamefont {Romero}},\ and\ \bibinfo {author} {\bibfnamefont {A.~C.}\
  \bibnamefont {Garcia-Castro}},\ }\bibfield  {title} {\bibinfo {title}
  {Anomalous hall conductivity control in ${\mathrm{mn}}_{3}\mathrm{NiN}$
  antiperovskite by epitaxial strain along the kagome plane},\ }\href
  {https://doi.org/10.1103/PhysRevB.106.195113} {\bibfield  {journal} {\bibinfo
   {journal} {Phys. Rev. B}\ }\textbf {\bibinfo {volume} {106}},\ \bibinfo
  {pages} {195113} (\bibinfo {year} {2022})}\BibitemShut {NoStop}%
\bibitem [{\citenamefont {Flórez-Gómez}\ \emph {et~al.}(2022)\citenamefont
  {Flórez-Gómez}, \citenamefont {Ibarra-Hernández},\ and\ \citenamefont
  {Garcia-Castro}}]{FLOREZGOMEZ2022169813}%
  \BibitemOpen
  \bibfield  {author} {\bibinfo {author} {\bibfnamefont {L.}~\bibnamefont
  {Flórez-Gómez}}, \bibinfo {author} {\bibfnamefont {W.}~\bibnamefont
  {Ibarra-Hernández}},\ and\ \bibinfo {author} {\bibfnamefont
  {A.}~\bibnamefont {Garcia-Castro}},\ }\bibfield  {title} {\bibinfo {title}
  {Lattice dynamics and spin–phonon coupling in the noncollinear
  antiferromagnetic antiperovskite mn3nin},\ }\href
  {https://doi.org/https://doi.org/10.1016/j.jmmm.2022.169813} {\bibfield
  {journal} {\bibinfo  {journal} {Journal of Magnetism and Magnetic Materials}\
  }\textbf {\bibinfo {volume} {562}},\ \bibinfo {pages} {169813} (\bibinfo
  {year} {2022})}\BibitemShut {NoStop}%
\bibitem [{\citenamefont {Duran-Pinilla}\ \emph {et~al.}(2022)\citenamefont
  {Duran-Pinilla}, \citenamefont {Romero},\ and\ \citenamefont
  {Garcia-Castro}}]{PhysRevMaterials.6.125003}%
  \BibitemOpen
  \bibfield  {author} {\bibinfo {author} {\bibfnamefont {J.~M.}\ \bibnamefont
  {Duran-Pinilla}}, \bibinfo {author} {\bibfnamefont {A.~H.}\ \bibnamefont
  {Romero}},\ and\ \bibinfo {author} {\bibfnamefont {A.~C.}\ \bibnamefont
  {Garcia-Castro}},\ }\bibfield  {title} {\bibinfo {title} {Chiral magnetism,
  lattice dynamics, and anomalous hall conductivity in
  ${\mathrm{v}}_{3}\mathrm{AuN}$ antiferromagnetic antiperovskite},\ }\href
  {https://doi.org/10.1103/PhysRevMaterials.6.125003} {\bibfield  {journal}
  {\bibinfo  {journal} {Phys. Rev. Mater.}\ }\textbf {\bibinfo {volume} {6}},\
  \bibinfo {pages} {125003} (\bibinfo {year} {2022})}\BibitemShut {NoStop}%
\bibitem [{\citenamefont {Ghimire}\ and\ \citenamefont
  {Mazin}(2020)}]{NatMat-2020-1}%
  \BibitemOpen
  \bibfield  {author} {\bibinfo {author} {\bibfnamefont {N.~J.}\ \bibnamefont
  {Ghimire}}\ and\ \bibinfo {author} {\bibfnamefont {I.~I.}\ \bibnamefont
  {Mazin}},\ }\bibfield  {title} {\bibinfo {title} {Topology and correlations
  on the kagome lattice},\ }\href {https://doi.org/10.1038/s41563-019-0589-8}
  {\bibfield  {journal} {\bibinfo  {journal} {Nature Materials}\ }\textbf
  {\bibinfo {volume} {19}},\ \bibinfo {pages} {137} (\bibinfo {year}
  {2020})}\BibitemShut {NoStop}%
\bibitem [{\citenamefont {Luo}\ \emph {et~al.}(2022)\citenamefont {Luo},
  \citenamefont {Zhao}, \citenamefont {Zhou}, \citenamefont {Yang},
  \citenamefont {Fang}, \citenamefont {Yang}, \citenamefont {Gao},
  \citenamefont {Zhou},\ and\ \citenamefont {Zheng}}]{AV3Sb5-NPJ-2022}%
  \BibitemOpen
  \bibfield  {author} {\bibinfo {author} {\bibfnamefont {J.}~\bibnamefont
  {Luo}}, \bibinfo {author} {\bibfnamefont {Z.}~\bibnamefont {Zhao}}, \bibinfo
  {author} {\bibfnamefont {Y.~Z.}\ \bibnamefont {Zhou}}, \bibinfo {author}
  {\bibfnamefont {J.}~\bibnamefont {Yang}}, \bibinfo {author} {\bibfnamefont
  {A.~F.}\ \bibnamefont {Fang}}, \bibinfo {author} {\bibfnamefont {H.~T.}\
  \bibnamefont {Yang}}, \bibinfo {author} {\bibfnamefont {H.~J.}\ \bibnamefont
  {Gao}}, \bibinfo {author} {\bibfnamefont {R.}~\bibnamefont {Zhou}},\ and\
  \bibinfo {author} {\bibfnamefont {G.-q.}\ \bibnamefont {Zheng}},\ }\bibfield
  {title} {\bibinfo {title} {Possible star-of-david pattern charge density wave
  with additional modulation in the kagome superconductor csv3sb5},\ }\href
  {https://doi.org/10.1038/s41535-022-00437-7} {\bibfield  {journal} {\bibinfo
  {journal} {npj Quantum Materials}\ }\textbf {\bibinfo {volume} {7}},\
  \bibinfo {pages} {30} (\bibinfo {year} {2022})}\BibitemShut {NoStop}%
\bibitem [{\citenamefont {Yin}\ \emph {et~al.}(2022)\citenamefont {Yin},
  \citenamefont {Lian},\ and\ \citenamefont {Hasan}}]{Kagome-nature-2023}%
  \BibitemOpen
  \bibfield  {author} {\bibinfo {author} {\bibfnamefont {J.-X.}\ \bibnamefont
  {Yin}}, \bibinfo {author} {\bibfnamefont {B.}~\bibnamefont {Lian}},\ and\
  \bibinfo {author} {\bibfnamefont {M.~Z.}\ \bibnamefont {Hasan}},\ }\bibfield
  {title} {\bibinfo {title} {Topological kagome magnets and superconductors},\
  }\href {https://doi.org/10.1038/s41586-022-05516-0} {\bibfield  {journal}
  {\bibinfo  {journal} {Nature}\ }\textbf {\bibinfo {volume} {612}},\ \bibinfo
  {pages} {647} (\bibinfo {year} {2022})}\BibitemShut {NoStop}%
\bibitem [{\citenamefont {Ortiz}\ \emph {et~al.}(2019)\citenamefont {Ortiz},
  \citenamefont {Gomes}, \citenamefont {Morey}, \citenamefont {Winiarski},
  \citenamefont {Bordelon}, \citenamefont {Mangum}, \citenamefont {Oswald},
  \citenamefont {Rodriguez-Rivera}, \citenamefont {Neilson}, \citenamefont
  {Wilson}, \citenamefont {Ertekin}, \citenamefont {McQueen},\ and\
  \citenamefont {Toberer}}]{PhysRevMaterials.3.094407}%
  \BibitemOpen
  \bibfield  {author} {\bibinfo {author} {\bibfnamefont {B.~R.}\ \bibnamefont
  {Ortiz}}, \bibinfo {author} {\bibfnamefont {L.~C.}\ \bibnamefont {Gomes}},
  \bibinfo {author} {\bibfnamefont {J.~R.}\ \bibnamefont {Morey}}, \bibinfo
  {author} {\bibfnamefont {M.}~\bibnamefont {Winiarski}}, \bibinfo {author}
  {\bibfnamefont {M.}~\bibnamefont {Bordelon}}, \bibinfo {author}
  {\bibfnamefont {J.~S.}\ \bibnamefont {Mangum}}, \bibinfo {author}
  {\bibfnamefont {I.~W.~H.}\ \bibnamefont {Oswald}}, \bibinfo {author}
  {\bibfnamefont {J.~A.}\ \bibnamefont {Rodriguez-Rivera}}, \bibinfo {author}
  {\bibfnamefont {J.~R.}\ \bibnamefont {Neilson}}, \bibinfo {author}
  {\bibfnamefont {S.~D.}\ \bibnamefont {Wilson}}, \bibinfo {author}
  {\bibfnamefont {E.}~\bibnamefont {Ertekin}}, \bibinfo {author} {\bibfnamefont
  {T.~M.}\ \bibnamefont {McQueen}},\ and\ \bibinfo {author} {\bibfnamefont
  {E.~S.}\ \bibnamefont {Toberer}},\ }\bibfield  {title} {\bibinfo {title} {New
  kagome prototype materials: discovery of
  ${\mathrm{kv}}_{3}{\mathrm{sb}}_{5},{\mathrm{rbv}}_{3}{\mathrm{sb}}_{5}$, and
  ${\mathrm{csv}}_{3}{\mathrm{sb}}_{5}$},\ }\href
  {https://doi.org/10.1103/PhysRevMaterials.3.094407} {\bibfield  {journal}
  {\bibinfo  {journal} {Phys. Rev. Mater.}\ }\textbf {\bibinfo {volume} {3}},\
  \bibinfo {pages} {094407} (\bibinfo {year} {2019})}\BibitemShut {NoStop}%
\bibitem [{\citenamefont {Subires}\ \emph {et~al.}(2023)\citenamefont
  {Subires}, \citenamefont {Korshunov}, \citenamefont {Said}, \citenamefont
  {S{\'a}nchez}, \citenamefont {Ortiz}, \citenamefont {Wilson}, \citenamefont
  {Bosak},\ and\ \citenamefont {Blanco-Canosa}}]{NatComm-2023-1}%
  \BibitemOpen
  \bibfield  {author} {\bibinfo {author} {\bibfnamefont {D.}~\bibnamefont
  {Subires}}, \bibinfo {author} {\bibfnamefont {A.}~\bibnamefont {Korshunov}},
  \bibinfo {author} {\bibfnamefont {A.~H.}\ \bibnamefont {Said}}, \bibinfo
  {author} {\bibfnamefont {L.}~\bibnamefont {S{\'a}nchez}}, \bibinfo {author}
  {\bibfnamefont {B.~R.}\ \bibnamefont {Ortiz}}, \bibinfo {author}
  {\bibfnamefont {S.~D.}\ \bibnamefont {Wilson}}, \bibinfo {author}
  {\bibfnamefont {A.}~\bibnamefont {Bosak}},\ and\ \bibinfo {author}
  {\bibfnamefont {S.}~\bibnamefont {Blanco-Canosa}},\ }\bibfield  {title}
  {\bibinfo {title} {Order-disorder charge density wave instability in the
  kagome metal (cs,rb)v3sb5},\ }\href
  {https://doi.org/10.1038/s41467-023-36668-w} {\bibfield  {journal} {\bibinfo
  {journal} {Nature Communications}\ }\textbf {\bibinfo {volume} {14}},\
  \bibinfo {pages} {1015} (\bibinfo {year} {2023})}\BibitemShut {NoStop}%
\bibitem [{\citenamefont {Uykur}\ \emph {et~al.}(2022)\citenamefont {Uykur},
  \citenamefont {Ortiz}, \citenamefont {Wilson}, \citenamefont {Dressel},\ and\
  \citenamefont {Tsirlin}}]{SciRep-2022-1}%
  \BibitemOpen
  \bibfield  {author} {\bibinfo {author} {\bibfnamefont {E.}~\bibnamefont
  {Uykur}}, \bibinfo {author} {\bibfnamefont {B.~R.}\ \bibnamefont {Ortiz}},
  \bibinfo {author} {\bibfnamefont {S.~D.}\ \bibnamefont {Wilson}}, \bibinfo
  {author} {\bibfnamefont {M.}~\bibnamefont {Dressel}},\ and\ \bibinfo {author}
  {\bibfnamefont {A.~A.}\ \bibnamefont {Tsirlin}},\ }\bibfield  {title}
  {\bibinfo {title} {Optical detection of the density-wave instability in the
  kagome metal kv3sb5},\ }\href {https://doi.org/10.1038/s41535-021-00420-8}
  {\bibfield  {journal} {\bibinfo  {journal} {npj Quantum Materials}\ }\textbf
  {\bibinfo {volume} {7}},\ \bibinfo {pages} {16} (\bibinfo {year}
  {2022})}\BibitemShut {NoStop}%
\bibitem [{\citenamefont {Ritz}\ \emph
  {et~al.}(2023{\natexlab{a}})\citenamefont {Ritz}, \citenamefont {Fernandes},\
  and\ \citenamefont {Birol}}]{PhysRevB.107.205131}%
  \BibitemOpen
  \bibfield  {author} {\bibinfo {author} {\bibfnamefont {E.~T.}\ \bibnamefont
  {Ritz}}, \bibinfo {author} {\bibfnamefont {R.~M.}\ \bibnamefont
  {Fernandes}},\ and\ \bibinfo {author} {\bibfnamefont {T.}~\bibnamefont
  {Birol}},\ }\bibfield  {title} {\bibinfo {title} {Impact of sb degrees of
  freedom on the charge density wave phase diagram of the kagome metal
  ${\mathrm{csv}}_{3}{\mathrm{sb}}_{5}$},\ }\href
  {https://doi.org/10.1103/PhysRevB.107.205131} {\bibfield  {journal} {\bibinfo
   {journal} {Phys. Rev. B}\ }\textbf {\bibinfo {volume} {107}},\ \bibinfo
  {pages} {205131} (\bibinfo {year} {2023}{\natexlab{a}})}\BibitemShut
  {NoStop}%
\bibitem [{\citenamefont {Ritz}\ \emph
  {et~al.}(2023{\natexlab{b}})\citenamefont {Ritz}, \citenamefont {Røising},
  \citenamefont {Christensen}, \citenamefont {Birol}, \citenamefont
  {Andersen},\ and\ \citenamefont {Fernandes}}]{ritz2023superconductivity}%
  \BibitemOpen
  \bibfield  {author} {\bibinfo {author} {\bibfnamefont {E.}~\bibnamefont
  {Ritz}}, \bibinfo {author} {\bibfnamefont {H.~S.}\ \bibnamefont {Røising}},
  \bibinfo {author} {\bibfnamefont {M.~H.}\ \bibnamefont {Christensen}},
  \bibinfo {author} {\bibfnamefont {T.}~\bibnamefont {Birol}}, \bibinfo
  {author} {\bibfnamefont {B.~M.}\ \bibnamefont {Andersen}},\ and\ \bibinfo
  {author} {\bibfnamefont {R.~M.}\ \bibnamefont {Fernandes}},\ }\href@noop {}
  {\bibinfo {title} {Superconductivity from orbital-selective electron-phonon
  coupling in $a\mathrm{V}_3\mathrm{Sb}_5$}} (\bibinfo {year}
  {2023}{\natexlab{b}}),\ \Eprint {https://arxiv.org/abs/2304.14822}
  {arXiv:2304.14822 [cond-mat.supr-con]} \BibitemShut {NoStop}%
\bibitem [{\citenamefont {Chen}\ \emph {et~al.}(2014)\citenamefont {Chen},
  \citenamefont {Niu},\ and\ \citenamefont
  {MacDonald}}]{PhysRevLett.112.017205}%
  \BibitemOpen
  \bibfield  {author} {\bibinfo {author} {\bibfnamefont {H.}~\bibnamefont
  {Chen}}, \bibinfo {author} {\bibfnamefont {Q.}~\bibnamefont {Niu}},\ and\
  \bibinfo {author} {\bibfnamefont {A.~H.}\ \bibnamefont {MacDonald}},\
  }\bibfield  {title} {\bibinfo {title} {Anomalous hall effect arising from
  noncollinear antiferromagnetism},\ }\href
  {https://doi.org/10.1103/PhysRevLett.112.017205} {\bibfield  {journal}
  {\bibinfo  {journal} {Phys. Rev. Lett.}\ }\textbf {\bibinfo {volume} {112}},\
  \bibinfo {pages} {017205} (\bibinfo {year} {2014})}\BibitemShut {NoStop}%
\bibitem [{\citenamefont {{\v S}mejkal}\ \emph {et~al.}(2022)\citenamefont {{\v
  S}mejkal}, \citenamefont {MacDonald}, \citenamefont {Sinova}, \citenamefont
  {Nakatsuji},\ and\ \citenamefont {Jungwirth}}]{AHC-AFM-NatRev-2022}%
  \BibitemOpen
  \bibfield  {author} {\bibinfo {author} {\bibfnamefont {L.}~\bibnamefont {{\v
  S}mejkal}}, \bibinfo {author} {\bibfnamefont {A.~H.}\ \bibnamefont
  {MacDonald}}, \bibinfo {author} {\bibfnamefont {J.}~\bibnamefont {Sinova}},
  \bibinfo {author} {\bibfnamefont {S.}~\bibnamefont {Nakatsuji}},\ and\
  \bibinfo {author} {\bibfnamefont {T.}~\bibnamefont {Jungwirth}},\ }\bibfield
  {title} {\bibinfo {title} {Anomalous hall antiferromagnets},\ }\href
  {https://doi.org/10.1038/s41578-022-00430-3} {\bibfield  {journal} {\bibinfo
  {journal} {Nature Reviews Materials}\ }\textbf {\bibinfo {volume} {7}},\
  \bibinfo {pages} {482} (\bibinfo {year} {2022})}\BibitemShut {NoStop}%
\bibitem [{\citenamefont {Wang}\ \emph {et~al.}(2007)\citenamefont {Wang},
  \citenamefont {Vanderbilt}, \citenamefont {Yates},\ and\ \citenamefont
  {Souza}}]{Souza2007}%
  \BibitemOpen
  \bibfield  {author} {\bibinfo {author} {\bibfnamefont {X.}~\bibnamefont
  {Wang}}, \bibinfo {author} {\bibfnamefont {D.}~\bibnamefont {Vanderbilt}},
  \bibinfo {author} {\bibfnamefont {J.~R.}\ \bibnamefont {Yates}},\ and\
  \bibinfo {author} {\bibfnamefont {I.}~\bibnamefont {Souza}},\ }\bibfield
  {title} {\bibinfo {title} {Fermi-surface calculation of the anomalous hall
  conductivity},\ }\href {https://doi.org/10.1103/PhysRevB.76.195109}
  {\bibfield  {journal} {\bibinfo  {journal} {Phys. Rev. B}\ }\textbf {\bibinfo
  {volume} {76}},\ \bibinfo {pages} {195109} (\bibinfo {year}
  {2007})}\BibitemShut {NoStop}%
\bibitem [{\citenamefont {Wang}\ \emph {et~al.}(2006)\citenamefont {Wang},
  \citenamefont {Yates}, \citenamefont {Souza},\ and\ \citenamefont
  {Vanderbilt}}]{DV2006}%
  \BibitemOpen
  \bibfield  {author} {\bibinfo {author} {\bibfnamefont {X.}~\bibnamefont
  {Wang}}, \bibinfo {author} {\bibfnamefont {J.~R.}\ \bibnamefont {Yates}},
  \bibinfo {author} {\bibfnamefont {I.}~\bibnamefont {Souza}},\ and\ \bibinfo
  {author} {\bibfnamefont {D.}~\bibnamefont {Vanderbilt}},\ }\bibfield  {title}
  {\bibinfo {title} {Ab initio calculation of the anomalous hall conductivity
  by wannier interpolation},\ }\href
  {https://doi.org/10.1103/PhysRevB.74.195118} {\bibfield  {journal} {\bibinfo
  {journal} {Phys. Rev. B}\ }\textbf {\bibinfo {volume} {74}},\ \bibinfo
  {pages} {195118} (\bibinfo {year} {2006})}\BibitemShut {NoStop}%
\bibitem [{\citenamefont {Singh}\ \emph {et~al.}(2020)\citenamefont {Singh},
  \citenamefont {Kim}, \citenamefont {Rabe},\ and\ \citenamefont
  {Vanderbilt}}]{SS2020}%
  \BibitemOpen
  \bibfield  {author} {\bibinfo {author} {\bibfnamefont {S.}~\bibnamefont
  {Singh}}, \bibinfo {author} {\bibfnamefont {J.}~\bibnamefont {Kim}}, \bibinfo
  {author} {\bibfnamefont {K.~M.}\ \bibnamefont {Rabe}},\ and\ \bibinfo
  {author} {\bibfnamefont {D.}~\bibnamefont {Vanderbilt}},\ }\bibfield  {title}
  {\bibinfo {title} {Engineering weyl phases and nonlinear hall effects in
  ${\mathrm{t}}_{d}$-${\mathrm{mote}}_{2}$},\ }\href
  {https://doi.org/10.1103/PhysRevLett.125.046402} {\bibfield  {journal}
  {\bibinfo  {journal} {Phys. Rev. Lett.}\ }\textbf {\bibinfo {volume} {125}},\
  \bibinfo {pages} {046402} (\bibinfo {year} {2020})}\BibitemShut {NoStop}%
\bibitem [{\citenamefont {Liu}\ \emph {et~al.}(2018{\natexlab{b}})\citenamefont
  {Liu}, \citenamefont {Sun}, \citenamefont {Kumar}, \citenamefont {Muechler},
  \citenamefont {Sun}, \citenamefont {Jiao}, \citenamefont {Yang},
  \citenamefont {Liu}, \citenamefont {Liang}, \citenamefont {Xu}, \citenamefont
  {Kroder}, \citenamefont {S{\"{u}}{\ss}}, \citenamefont {Borrmann},
  \citenamefont {Shekhar}, \citenamefont {Wang}, \citenamefont {Xi},
  \citenamefont {Wang}, \citenamefont {Schnelle}, \citenamefont {Wirth},
  \citenamefont {Chen}, \citenamefont {Goennenwein},\ and\ \citenamefont
  {Felser}}]{Liu2018}%
  \BibitemOpen
  \bibfield  {author} {\bibinfo {author} {\bibfnamefont {E.}~\bibnamefont
  {Liu}}, \bibinfo {author} {\bibfnamefont {Y.}~\bibnamefont {Sun}}, \bibinfo
  {author} {\bibfnamefont {N.}~\bibnamefont {Kumar}}, \bibinfo {author}
  {\bibfnamefont {L.}~\bibnamefont {Muechler}}, \bibinfo {author}
  {\bibfnamefont {A.}~\bibnamefont {Sun}}, \bibinfo {author} {\bibfnamefont
  {L.}~\bibnamefont {Jiao}}, \bibinfo {author} {\bibfnamefont {S.~Y.}\
  \bibnamefont {Yang}}, \bibinfo {author} {\bibfnamefont {D.}~\bibnamefont
  {Liu}}, \bibinfo {author} {\bibfnamefont {A.}~\bibnamefont {Liang}}, \bibinfo
  {author} {\bibfnamefont {Q.}~\bibnamefont {Xu}}, \bibinfo {author}
  {\bibfnamefont {J.}~\bibnamefont {Kroder}}, \bibinfo {author} {\bibfnamefont
  {V.}~\bibnamefont {S{\"{u}}{\ss}}}, \bibinfo {author} {\bibfnamefont
  {H.}~\bibnamefont {Borrmann}}, \bibinfo {author} {\bibfnamefont
  {C.}~\bibnamefont {Shekhar}}, \bibinfo {author} {\bibfnamefont
  {Z.}~\bibnamefont {Wang}}, \bibinfo {author} {\bibfnamefont {C.}~\bibnamefont
  {Xi}}, \bibinfo {author} {\bibfnamefont {W.}~\bibnamefont {Wang}}, \bibinfo
  {author} {\bibfnamefont {W.}~\bibnamefont {Schnelle}}, \bibinfo {author}
  {\bibfnamefont {S.}~\bibnamefont {Wirth}}, \bibinfo {author} {\bibfnamefont
  {Y.}~\bibnamefont {Chen}}, \bibinfo {author} {\bibfnamefont {S.~T.}\
  \bibnamefont {Goennenwein}},\ and\ \bibinfo {author} {\bibfnamefont
  {C.}~\bibnamefont {Felser}},\ }\bibfield  {title} {\bibinfo {title} {{Giant
  anomalous Hall effect in a ferromagnetic kagome-lattice semimetal}},\ }\href
  {https://doi.org/10.1038/s41567-018-0234-5} {\bibfield  {journal} {\bibinfo
  {journal} {Nature Physics}\ }\textbf {\bibinfo {volume} {14}},\ \bibinfo
  {pages} {1125} (\bibinfo {year} {2018}{\natexlab{b}})}\BibitemShut {NoStop}%
\bibitem [{\citenamefont {Tsai}\ \emph {et~al.}(2020)\citenamefont {Tsai},
  \citenamefont {Higo}, \citenamefont {Kondou}, \citenamefont {Nomoto},
  \citenamefont {Sakai}, \citenamefont {Kobayashi}, \citenamefont {Nakano},
  \citenamefont {Yakushiji}, \citenamefont {Arita}, \citenamefont {Miwa},
  \citenamefont {Otani},\ and\ \citenamefont {Nakatsuji}}]{Tsai2020}%
  \BibitemOpen
  \bibfield  {author} {\bibinfo {author} {\bibfnamefont {H.}~\bibnamefont
  {Tsai}}, \bibinfo {author} {\bibfnamefont {T.}~\bibnamefont {Higo}}, \bibinfo
  {author} {\bibfnamefont {K.}~\bibnamefont {Kondou}}, \bibinfo {author}
  {\bibfnamefont {T.}~\bibnamefont {Nomoto}}, \bibinfo {author} {\bibfnamefont
  {A.}~\bibnamefont {Sakai}}, \bibinfo {author} {\bibfnamefont
  {A.}~\bibnamefont {Kobayashi}}, \bibinfo {author} {\bibfnamefont
  {T.}~\bibnamefont {Nakano}}, \bibinfo {author} {\bibfnamefont
  {K.}~\bibnamefont {Yakushiji}}, \bibinfo {author} {\bibfnamefont
  {R.}~\bibnamefont {Arita}}, \bibinfo {author} {\bibfnamefont
  {S.}~\bibnamefont {Miwa}}, \bibinfo {author} {\bibfnamefont {Y.}~\bibnamefont
  {Otani}},\ and\ \bibinfo {author} {\bibfnamefont {S.}~\bibnamefont
  {Nakatsuji}},\ }\bibfield  {title} {\bibinfo {title} {{Electrical
  manipulation of a topological antiferromagnetic state}},\ }\href
  {https://doi.org/10.1038/s41586-020-2211-2} {\bibfield  {journal} {\bibinfo
  {journal} {Nature}\ ,\ \bibinfo {pages} {1}} (\bibinfo {year}
  {2020})}\BibitemShut {NoStop}%
\bibitem [{\citenamefont {Qin}\ \emph {et~al.}(2019)\citenamefont {Qin},
  \citenamefont {Yan}, \citenamefont {Wang}, \citenamefont {Feng},
  \citenamefont {Guo}, \citenamefont {Zhou}, \citenamefont {Wu}, \citenamefont
  {Zhang}, \citenamefont {Leng}, \citenamefont {Chen},\ and\ \citenamefont
  {Liu}}]{Qin2019}%
  \BibitemOpen
  \bibfield  {author} {\bibinfo {author} {\bibfnamefont {P.~X.}\ \bibnamefont
  {Qin}}, \bibinfo {author} {\bibfnamefont {H.}~\bibnamefont {Yan}}, \bibinfo
  {author} {\bibfnamefont {X.~N.}\ \bibnamefont {Wang}}, \bibinfo {author}
  {\bibfnamefont {Z.~X.}\ \bibnamefont {Feng}}, \bibinfo {author}
  {\bibfnamefont {H.~X.}\ \bibnamefont {Guo}}, \bibinfo {author} {\bibfnamefont
  {X.~R.}\ \bibnamefont {Zhou}}, \bibinfo {author} {\bibfnamefont {H.~J.}\
  \bibnamefont {Wu}}, \bibinfo {author} {\bibfnamefont {X.}~\bibnamefont
  {Zhang}}, \bibinfo {author} {\bibfnamefont {Z.~G.~G.}\ \bibnamefont {Leng}},
  \bibinfo {author} {\bibfnamefont {H.~Y.}\ \bibnamefont {Chen}},\ and\
  \bibinfo {author} {\bibfnamefont {Z.~Q.}\ \bibnamefont {Liu}},\ }\bibfield
  {title} {\bibinfo {title} {{Noncollinear spintronics and electric-field
  control: a review}},\ }\bibfield  {journal} {\bibinfo  {journal} {Rare
  Metals}\ }\href {https://doi.org/10.1007/s12598-019-01352-w}
  {10.1007/s12598-019-01352-w} (\bibinfo {year} {2019})\BibitemShut {NoStop}%
\bibitem [{\citenamefont {Bai}\ \emph {et~al.}(2022)\citenamefont {Bai},
  \citenamefont {Zhang}, \citenamefont {Han}, \citenamefont {Zhou},
  \citenamefont {Pan},\ and\ \citenamefont {Song}}]{doi:10.1063/5.0101981}%
  \BibitemOpen
  \bibfield  {author} {\bibinfo {author} {\bibfnamefont {H.}~\bibnamefont
  {Bai}}, \bibinfo {author} {\bibfnamefont {Y.~C.}\ \bibnamefont {Zhang}},
  \bibinfo {author} {\bibfnamefont {L.}~\bibnamefont {Han}}, \bibinfo {author}
  {\bibfnamefont {Y.~J.}\ \bibnamefont {Zhou}}, \bibinfo {author}
  {\bibfnamefont {F.}~\bibnamefont {Pan}},\ and\ \bibinfo {author}
  {\bibfnamefont {C.}~\bibnamefont {Song}},\ }\bibfield  {title} {\bibinfo
  {title} {Antiferromagnetism: An efficient and controllable spin source},\
  }\href {https://doi.org/10.1063/5.0101981} {\bibfield  {journal} {\bibinfo
  {journal} {Applied Physics Reviews}\ }\textbf {\bibinfo {volume} {9}},\
  \bibinfo {pages} {041316} (\bibinfo {year} {2022})},\ \Eprint
  {https://arxiv.org/abs/https://doi.org/10.1063/5.0101981}
  {https://doi.org/10.1063/5.0101981} \BibitemShut {NoStop}%
\bibitem [{\citenamefont {Jiang}\ \emph {et~al.}(2022)\citenamefont {Jiang},
  \citenamefont {Yu}, \citenamefont {Wang}, \citenamefont {Lu}, \citenamefont
  {Meng}, \citenamefont {Jiang},\ and\ \citenamefont {Liu}}]{Jiang_2022}%
  \BibitemOpen
  \bibfield  {author} {\bibinfo {author} {\bibfnamefont {Y.}~\bibnamefont
  {Jiang}}, \bibinfo {author} {\bibfnamefont {Z.}~\bibnamefont {Yu}}, \bibinfo
  {author} {\bibfnamefont {Y.}~\bibnamefont {Wang}}, \bibinfo {author}
  {\bibfnamefont {T.}~\bibnamefont {Lu}}, \bibinfo {author} {\bibfnamefont
  {S.}~\bibnamefont {Meng}}, \bibinfo {author} {\bibfnamefont {K.}~\bibnamefont
  {Jiang}},\ and\ \bibinfo {author} {\bibfnamefont {M.}~\bibnamefont {Liu}},\
  }\bibfield  {title} {\bibinfo {title} {Screening promising csv3sb5-like
  kagome materials from systematic first-principles evaluation},\ }\href
  {https://doi.org/10.1088/0256-307X/39/4/047402} {\bibfield  {journal}
  {\bibinfo  {journal} {Chinese Physics Letters}\ }\textbf {\bibinfo {volume}
  {39}},\ \bibinfo {pages} {047402} (\bibinfo {year} {2022})}\BibitemShut
  {NoStop}%
\bibitem [{\citenamefont {Hohenberg}\ and\ \citenamefont
  {Kohn}(1964)}]{PhysRev.136.B864}%
  \BibitemOpen
  \bibfield  {author} {\bibinfo {author} {\bibfnamefont {P.}~\bibnamefont
  {Hohenberg}}\ and\ \bibinfo {author} {\bibfnamefont {W.}~\bibnamefont
  {Kohn}},\ }\bibfield  {title} {\bibinfo {title} {Inhomogeneous electron
  gas},\ }\href {https://doi.org/10.1103/PhysRev.136.B864} {\bibfield
  {journal} {\bibinfo  {journal} {Phys. Rev.}\ }\textbf {\bibinfo {volume}
  {136}},\ \bibinfo {pages} {B864} (\bibinfo {year} {1964})}\BibitemShut
  {NoStop}%
\bibitem [{\citenamefont {Kohn}\ and\ \citenamefont
  {Sham}(1965)}]{PhysRev.140.A1133}%
  \BibitemOpen
  \bibfield  {author} {\bibinfo {author} {\bibfnamefont {W.}~\bibnamefont
  {Kohn}}\ and\ \bibinfo {author} {\bibfnamefont {L.~J.}\ \bibnamefont
  {Sham}},\ }\bibfield  {title} {\bibinfo {title} {Self-consistent equations
  including exchange and correlation effects},\ }\href
  {https://doi.org/10.1103/PhysRev.140.A1133} {\bibfield  {journal} {\bibinfo
  {journal} {Phys. Rev.}\ }\textbf {\bibinfo {volume} {140}},\ \bibinfo {pages}
  {A1133} (\bibinfo {year} {1965})}\BibitemShut {NoStop}%
\bibitem [{\citenamefont {Bl\"ochl}(1994)}]{Blochl1994}%
  \BibitemOpen
  \bibfield  {author} {\bibinfo {author} {\bibfnamefont {P.~E.}\ \bibnamefont
  {Bl\"ochl}},\ }\bibfield  {title} {\bibinfo {title} {Projector augmented-wave
  method},\ }\href {https://doi.org/10.1103/PhysRevB.50.17953} {\bibfield
  {journal} {\bibinfo  {journal} {Phys. Rev. B}\ }\textbf {\bibinfo {volume}
  {50}},\ \bibinfo {pages} {17953} (\bibinfo {year} {1994})}\BibitemShut
  {NoStop}%
\bibitem [{\citenamefont {Kresse}\ and\ \citenamefont
  {Furthm\"uller}(1996)}]{Kresse1996}%
  \BibitemOpen
  \bibfield  {author} {\bibinfo {author} {\bibfnamefont {G.}~\bibnamefont
  {Kresse}}\ and\ \bibinfo {author} {\bibfnamefont {J.}~\bibnamefont
  {Furthm\"uller}},\ }\bibfield  {title} {\bibinfo {title} {Efficient iterative
  schemes for \textit{ab initio} total-energy calculations using a plane-wave
  basis set},\ }\href {https://doi.org/10.1103/PhysRevB.54.11169} {\bibfield
  {journal} {\bibinfo  {journal} {Phys. Rev. B}\ }\textbf {\bibinfo {volume}
  {54}},\ \bibinfo {pages} {11169} (\bibinfo {year} {1996})}\BibitemShut
  {NoStop}%
\bibitem [{\citenamefont {Kresse}\ and\ \citenamefont
  {Joubert}(1999)}]{Kresse1999}%
  \BibitemOpen
  \bibfield  {author} {\bibinfo {author} {\bibfnamefont {G.}~\bibnamefont
  {Kresse}}\ and\ \bibinfo {author} {\bibfnamefont {D.}~\bibnamefont
  {Joubert}},\ }\bibfield  {title} {\bibinfo {title} {From ultrasoft
  pseudopotentials to the projector augmented-wave method},\ }\href
  {https://doi.org/10.1103/PhysRevB.59.1758} {\bibfield  {journal} {\bibinfo
  {journal} {Phys. Rev. B}\ }\textbf {\bibinfo {volume} {59}},\ \bibinfo
  {pages} {1758} (\bibinfo {year} {1999})}\BibitemShut {NoStop}%
\bibitem [{\citenamefont {Perdew}\ \emph {et~al.}(2008)\citenamefont {Perdew},
  \citenamefont {Ruzsinszky}, \citenamefont {Csonka}, \citenamefont {Vydrov},
  \citenamefont {Scuseria}, \citenamefont {Constantin}, \citenamefont {Zhou},\
  and\ \citenamefont {Burke}}]{Perdew2008}%
  \BibitemOpen
  \bibfield  {author} {\bibinfo {author} {\bibfnamefont {J.~P.}\ \bibnamefont
  {Perdew}}, \bibinfo {author} {\bibfnamefont {A.}~\bibnamefont {Ruzsinszky}},
  \bibinfo {author} {\bibfnamefont {G.~I.}\ \bibnamefont {Csonka}}, \bibinfo
  {author} {\bibfnamefont {O.~A.}\ \bibnamefont {Vydrov}}, \bibinfo {author}
  {\bibfnamefont {G.~E.}\ \bibnamefont {Scuseria}}, \bibinfo {author}
  {\bibfnamefont {L.~A.}\ \bibnamefont {Constantin}}, \bibinfo {author}
  {\bibfnamefont {X.}~\bibnamefont {Zhou}},\ and\ \bibinfo {author}
  {\bibfnamefont {K.}~\bibnamefont {Burke}},\ }\bibfield  {title} {\bibinfo
  {title} {Restoring the density-gradient expansion for exchange in solids and
  surfaces},\ }\href {https://doi.org/10.1103/PhysRevLett.100.136406}
  {\bibfield  {journal} {\bibinfo  {journal} {Phys. Rev. Lett.}\ }\textbf
  {\bibinfo {volume} {100}},\ \bibinfo {pages} {136406} (\bibinfo {year}
  {2008})}\BibitemShut {NoStop}%
\bibitem [{\citenamefont {Liechtenstein}\ \emph {et~al.}(1995)\citenamefont
  {Liechtenstein}, \citenamefont {Anisimov},\ and\ \citenamefont
  {Zaanen}}]{Liechtenstein1995}%
  \BibitemOpen
  \bibfield  {author} {\bibinfo {author} {\bibfnamefont {A.~I.}\ \bibnamefont
  {Liechtenstein}}, \bibinfo {author} {\bibfnamefont {V.~I.}\ \bibnamefont
  {Anisimov}},\ and\ \bibinfo {author} {\bibfnamefont {J.}~\bibnamefont
  {Zaanen}},\ }\bibfield  {title} {\bibinfo {title} {Density-functional theory
  and strong interactions: Orbital ordering in mott-hubbard insulators},\
  }\href {https://doi.org/10.1103/PhysRevB.52.R5467} {\bibfield  {journal}
  {\bibinfo  {journal} {Phys. Rev. B}\ }\textbf {\bibinfo {volume} {52}},\
  \bibinfo {pages} {R5467} (\bibinfo {year} {1995})}\BibitemShut {NoStop}%
\bibitem [{\citenamefont {Monkhorst}\ and\ \citenamefont
  {Pack}(1976)}]{PhysRevB.13.5188}%
  \BibitemOpen
  \bibfield  {author} {\bibinfo {author} {\bibfnamefont {H.~J.}\ \bibnamefont
  {Monkhorst}}\ and\ \bibinfo {author} {\bibfnamefont {J.~D.}\ \bibnamefont
  {Pack}},\ }\bibfield  {title} {\bibinfo {title} {Special points for
  brillouin-zone integrations},\ }\href
  {https://doi.org/10.1103/PhysRevB.13.5188} {\bibfield  {journal} {\bibinfo
  {journal} {Phys. Rev. B}\ }\textbf {\bibinfo {volume} {13}},\ \bibinfo
  {pages} {5188} (\bibinfo {year} {1976})}\BibitemShut {NoStop}%
\bibitem [{\citenamefont {Hobbs}\ \emph {et~al.}(2000)\citenamefont {Hobbs},
  \citenamefont {Kresse},\ and\ \citenamefont {Hafner}}]{Hobbs2000}%
  \BibitemOpen
  \bibfield  {author} {\bibinfo {author} {\bibfnamefont {D.}~\bibnamefont
  {Hobbs}}, \bibinfo {author} {\bibfnamefont {G.}~\bibnamefont {Kresse}},\ and\
  \bibinfo {author} {\bibfnamefont {J.}~\bibnamefont {Hafner}},\ }\bibfield
  {title} {\bibinfo {title} {Fully unconstrained noncollinear magnetism within
  the projector augmented-wave method},\ }\href
  {https://doi.org/10.1103/PhysRevB.62.11556} {\bibfield  {journal} {\bibinfo
  {journal} {Phys. Rev. B}\ }\textbf {\bibinfo {volume} {62}},\ \bibinfo
  {pages} {11556} (\bibinfo {year} {2000})}\BibitemShut {NoStop}%
\bibitem [{\citenamefont {Kunc}\ and\ \citenamefont
  {Martin}(1982)}]{PhysRevLett.48.406}%
  \BibitemOpen
  \bibfield  {author} {\bibinfo {author} {\bibfnamefont {K.}~\bibnamefont
  {Kunc}}\ and\ \bibinfo {author} {\bibfnamefont {R.~M.}\ \bibnamefont
  {Martin}},\ }\bibfield  {title} {\bibinfo {title} {Ab initio force constants
  of gaas: A new approach to calculation of phonons and dielectric
  properties},\ }\href {https://doi.org/10.1103/PhysRevLett.48.406} {\bibfield
  {journal} {\bibinfo  {journal} {Phys. Rev. Lett.}\ }\textbf {\bibinfo
  {volume} {48}},\ \bibinfo {pages} {406} (\bibinfo {year} {1982})}\BibitemShut
  {NoStop}%
\bibitem [{\citenamefont {Lam}\ \emph {et~al.}(1986)\citenamefont {Lam},
  \citenamefont {Dacorogna},\ and\ \citenamefont {Cohen}}]{PhysRevB.34.5065}%
  \BibitemOpen
  \bibfield  {author} {\bibinfo {author} {\bibfnamefont {P.~K.}\ \bibnamefont
  {Lam}}, \bibinfo {author} {\bibfnamefont {M.~M.}\ \bibnamefont {Dacorogna}},\
  and\ \bibinfo {author} {\bibfnamefont {M.~L.}\ \bibnamefont {Cohen}},\
  }\bibfield  {title} {\bibinfo {title} {Self-consistent calculation of
  electron-phonon couplings},\ }\href
  {https://doi.org/10.1103/PhysRevB.34.5065} {\bibfield  {journal} {\bibinfo
  {journal} {Phys. Rev. B}\ }\textbf {\bibinfo {volume} {34}},\ \bibinfo
  {pages} {5065} (\bibinfo {year} {1986})}\BibitemShut {NoStop}%
\bibitem [{\citenamefont {Togo}\ and\ \citenamefont {Tanaka}(2015)}]{phonopy}%
  \BibitemOpen
  \bibfield  {author} {\bibinfo {author} {\bibfnamefont {A.}~\bibnamefont
  {Togo}}\ and\ \bibinfo {author} {\bibfnamefont {I.}~\bibnamefont {Tanaka}},\
  }\bibfield  {title} {\bibinfo {title} {First principles phonon calculations
  in materials science},\ }\href@noop {} {\bibfield  {journal} {\bibinfo
  {journal} {Scr. Mater.}\ }\textbf {\bibinfo {volume} {108}},\ \bibinfo
  {pages} {1} (\bibinfo {year} {2015})}\BibitemShut {NoStop}%
\bibitem [{\citenamefont {Mostofi}\ \emph {et~al.}(2014)\citenamefont
  {Mostofi}, \citenamefont {Yates}, \citenamefont {Pizzi}, \citenamefont {Lee},
  \citenamefont {Souza}, \citenamefont {Vanderbilt},\ and\ \citenamefont
  {Marzari}}]{MOSTOFI20142309}%
  \BibitemOpen
  \bibfield  {author} {\bibinfo {author} {\bibfnamefont {A.~A.}\ \bibnamefont
  {Mostofi}}, \bibinfo {author} {\bibfnamefont {J.~R.}\ \bibnamefont {Yates}},
  \bibinfo {author} {\bibfnamefont {G.}~\bibnamefont {Pizzi}}, \bibinfo
  {author} {\bibfnamefont {Y.-S.}\ \bibnamefont {Lee}}, \bibinfo {author}
  {\bibfnamefont {I.}~\bibnamefont {Souza}}, \bibinfo {author} {\bibfnamefont
  {D.}~\bibnamefont {Vanderbilt}},\ and\ \bibinfo {author} {\bibfnamefont
  {N.}~\bibnamefont {Marzari}},\ }\bibfield  {title} {\bibinfo {title} {An
  updated version of wannier90: A tool for obtaining maximally-localised
  wannier functions},\ }\href
  {https://doi.org/https://doi.org/10.1016/j.cpc.2014.05.003} {\bibfield
  {journal} {\bibinfo  {journal} {Computer Physics Communications}\ }\textbf
  {\bibinfo {volume} {185}},\ \bibinfo {pages} {2309} (\bibinfo {year}
  {2014})}\BibitemShut {NoStop}%
\bibitem [{\citenamefont {Pizzi}\ \emph {et~al.}(2020)\citenamefont {Pizzi},
  \citenamefont {Vitale}, \citenamefont {Arita}, \citenamefont {Blügel},
  \citenamefont {Freimuth}, \citenamefont {G{\'{e}}ranton}, \citenamefont
  {Gibertini}, \citenamefont {Gresch}, \citenamefont {Johnson}, \citenamefont
  {Koretsune}, \citenamefont {Iba{\~{n}}ez-Azpiroz}, \citenamefont {Lee},
  \citenamefont {Lihm}, \citenamefont {Marchand}, \citenamefont {Marrazzo},
  \citenamefont {Mokrousov}, \citenamefont {Mustafa}, \citenamefont {Nohara},
  \citenamefont {Nomura}, \citenamefont {Paulatto}, \citenamefont
  {Ponc{\'{e}}}, \citenamefont {Ponweiser}, \citenamefont {Qiao}, \citenamefont
  {Thöle}, \citenamefont {Tsirkin}, \citenamefont {Wierzbowska}, \citenamefont
  {Marzari}, \citenamefont {Vanderbilt}, \citenamefont {Souza}, \citenamefont
  {Mostofi},\ and\ \citenamefont {Yates}}]{Pizzi_2020}%
  \BibitemOpen
  \bibfield  {author} {\bibinfo {author} {\bibfnamefont {G.}~\bibnamefont
  {Pizzi}}, \bibinfo {author} {\bibfnamefont {V.}~\bibnamefont {Vitale}},
  \bibinfo {author} {\bibfnamefont {R.}~\bibnamefont {Arita}}, \bibinfo
  {author} {\bibfnamefont {S.}~\bibnamefont {Blügel}}, \bibinfo {author}
  {\bibfnamefont {F.}~\bibnamefont {Freimuth}}, \bibinfo {author}
  {\bibfnamefont {G.}~\bibnamefont {G{\'{e}}ranton}}, \bibinfo {author}
  {\bibfnamefont {M.}~\bibnamefont {Gibertini}}, \bibinfo {author}
  {\bibfnamefont {D.}~\bibnamefont {Gresch}}, \bibinfo {author} {\bibfnamefont
  {C.}~\bibnamefont {Johnson}}, \bibinfo {author} {\bibfnamefont
  {T.}~\bibnamefont {Koretsune}}, \bibinfo {author} {\bibfnamefont
  {J.}~\bibnamefont {Iba{\~{n}}ez-Azpiroz}}, \bibinfo {author} {\bibfnamefont
  {H.}~\bibnamefont {Lee}}, \bibinfo {author} {\bibfnamefont {J.-M.}\
  \bibnamefont {Lihm}}, \bibinfo {author} {\bibfnamefont {D.}~\bibnamefont
  {Marchand}}, \bibinfo {author} {\bibfnamefont {A.}~\bibnamefont {Marrazzo}},
  \bibinfo {author} {\bibfnamefont {Y.}~\bibnamefont {Mokrousov}}, \bibinfo
  {author} {\bibfnamefont {J.~I.}\ \bibnamefont {Mustafa}}, \bibinfo {author}
  {\bibfnamefont {Y.}~\bibnamefont {Nohara}}, \bibinfo {author} {\bibfnamefont
  {Y.}~\bibnamefont {Nomura}}, \bibinfo {author} {\bibfnamefont
  {L.}~\bibnamefont {Paulatto}}, \bibinfo {author} {\bibfnamefont
  {S.}~\bibnamefont {Ponc{\'{e}}}}, \bibinfo {author} {\bibfnamefont
  {T.}~\bibnamefont {Ponweiser}}, \bibinfo {author} {\bibfnamefont
  {J.}~\bibnamefont {Qiao}}, \bibinfo {author} {\bibfnamefont {F.}~\bibnamefont
  {Thöle}}, \bibinfo {author} {\bibfnamefont {S.~S.}\ \bibnamefont {Tsirkin}},
  \bibinfo {author} {\bibfnamefont {M.}~\bibnamefont {Wierzbowska}}, \bibinfo
  {author} {\bibfnamefont {N.}~\bibnamefont {Marzari}}, \bibinfo {author}
  {\bibfnamefont {D.}~\bibnamefont {Vanderbilt}}, \bibinfo {author}
  {\bibfnamefont {I.}~\bibnamefont {Souza}}, \bibinfo {author} {\bibfnamefont
  {A.~A.}\ \bibnamefont {Mostofi}},\ and\ \bibinfo {author} {\bibfnamefont
  {J.~R.}\ \bibnamefont {Yates}},\ }\bibfield  {title} {\bibinfo {title}
  {Wannier90 as a community code: new features and applications},\ }\href
  {https://doi.org/10.1088/1361-648x/ab51ff} {\bibfield  {journal} {\bibinfo
  {journal} {Journal of Physics: Condensed Matter}\ }\textbf {\bibinfo {volume}
  {32}},\ \bibinfo {pages} {165902} (\bibinfo {year} {2020})}\BibitemShut
  {NoStop}%
\bibitem [{\citenamefont {Tsirkin}(2021)}]{wannierberri}%
  \BibitemOpen
  \bibfield  {author} {\bibinfo {author} {\bibfnamefont {S.~S.}\ \bibnamefont
  {Tsirkin}},\ }\bibfield  {title} {\bibinfo {title} {High performance wannier
  interpolation of berry curvature and related quantities with wannierberri
  code},\ }\href {https://doi.org/10.1038/s41524-021-00498-5} {\bibfield
  {journal} {\bibinfo  {journal} {npj Computational Materials}\ }\textbf
  {\bibinfo {volume} {7}},\ \bibinfo {pages} {33} (\bibinfo {year}
  {2021})}\BibitemShut {NoStop}%
\bibitem [{\citenamefont {Kawamura}(2019)}]{KAWAMURA2019197}%
  \BibitemOpen
  \bibfield  {author} {\bibinfo {author} {\bibfnamefont {M.}~\bibnamefont
  {Kawamura}},\ }\bibfield  {title} {\bibinfo {title} {Fermisurfer:
  Fermi-surface viewer providing multiple representation schemes},\ }\href
  {https://doi.org/https://doi.org/10.1016/j.cpc.2019.01.017} {\bibfield
  {journal} {\bibinfo  {journal} {Computer Physics Communications}\ }\textbf
  {\bibinfo {volume} {239}},\ \bibinfo {pages} {197} (\bibinfo {year}
  {2019})}\BibitemShut {NoStop}%
\bibitem [{\citenamefont {Momma}\ and\ \citenamefont {Izumi}(2011)}]{vesta}%
  \BibitemOpen
  \bibfield  {author} {\bibinfo {author} {\bibfnamefont {K.}~\bibnamefont
  {Momma}}\ and\ \bibinfo {author} {\bibfnamefont {F.}~\bibnamefont {Izumi}},\
  }\bibfield  {title} {\bibinfo {title} {{{\it VESTA3} for three-dimensional
  visualization of crystal, volumetric and morphology data}},\ }\href
  {https://doi.org/10.1107/S0021889811038970} {\bibfield  {journal} {\bibinfo
  {journal} {Journal of Applied Crystallography}\ }\textbf {\bibinfo {volume}
  {44}},\ \bibinfo {pages} {1272} (\bibinfo {year} {2011})}\BibitemShut
  {NoStop}%
\bibitem [{\citenamefont {Herath}\ \emph {et~al.}(2020)\citenamefont {Herath},
  \citenamefont {Tavadze}, \citenamefont {He}, \citenamefont {Bousquet},
  \citenamefont {Singh}, \citenamefont {Mu{\~n}oz},\ and\ \citenamefont
  {Romero}}]{HERATH2020107080}%
  \BibitemOpen
  \bibfield  {author} {\bibinfo {author} {\bibfnamefont {U.}~\bibnamefont
  {Herath}}, \bibinfo {author} {\bibfnamefont {P.}~\bibnamefont {Tavadze}},
  \bibinfo {author} {\bibfnamefont {X.}~\bibnamefont {He}}, \bibinfo {author}
  {\bibfnamefont {E.}~\bibnamefont {Bousquet}}, \bibinfo {author}
  {\bibfnamefont {S.}~\bibnamefont {Singh}}, \bibinfo {author} {\bibfnamefont
  {F.}~\bibnamefont {Mu{\~n}oz}},\ and\ \bibinfo {author} {\bibfnamefont
  {A.~H.}\ \bibnamefont {Romero}},\ }\bibfield  {title} {\bibinfo {title}
  {Pyprocar: A python library for electronic structure pre/post-processing},\
  }\href {https://doi.org/https://doi.org/10.1016/j.cpc.2019.107080} {\bibfield
   {journal} {\bibinfo  {journal} {Computer Physics Communications}\ }\textbf
  {\bibinfo {volume} {251}},\ \bibinfo {pages} {107080} (\bibinfo {year}
  {2020})}\BibitemShut {NoStop}%
\bibitem [{\citenamefont {Kawamura}\ and\ \citenamefont
  {Miyashita}(1984)}]{doi:10.1143/JPSJ.53.4138}%
  \BibitemOpen
  \bibfield  {author} {\bibinfo {author} {\bibfnamefont {H.}~\bibnamefont
  {Kawamura}}\ and\ \bibinfo {author} {\bibfnamefont {S.}~\bibnamefont
  {Miyashita}},\ }\bibfield  {title} {\bibinfo {title} {Phase transition of the
  two-dimensional heisenberg antiferromagnet on the triangular lattice},\
  }\href {https://doi.org/10.1143/JPSJ.53.4138} {\bibfield  {journal} {\bibinfo
   {journal} {Journal of the Physical Society of Japan}\ }\textbf {\bibinfo
  {volume} {53}},\ \bibinfo {pages} {4138} (\bibinfo {year} {1984})},\ \Eprint
  {https://arxiv.org/abs/https://doi.org/10.1143/JPSJ.53.4138}
  {https://doi.org/10.1143/JPSJ.53.4138} \BibitemShut {NoStop}%
\bibitem [{Note1()}]{Note1}%
  \BibitemOpen
  \bibinfo {note} {The magnetic vector chirality is defined as \protect \textbf
  {$\kappa $} = $\protect \frac {2}{3 \protect \sqrt {3}}\DOTSB \sum@ \slimits@
  _{i,j}$(\protect \textbf {S$_i$}$\times $\protect \textbf {S$_j$}) =
  $\protect \frac {2}{3 \protect \sqrt {3}}$(\protect \textbf {S$_1$}$\times
  $\protect \textbf {S$_2$}+\protect \textbf {S$_2$}$\times $\protect \textbf
  {S$_3$}+\protect \textbf {S$_3$}$\times $\protect \textbf {S$_1$}) for the
  kagome lattice where the $i$ and $j$ index run over the magnetic moments in
  the unit cell}\BibitemShut {NoStop}%
\bibitem [{\citenamefont {Ye}\ \emph {et~al.}(2018)\citenamefont {Ye},
  \citenamefont {Kang}, \citenamefont {Liu}, \citenamefont {von Cube},
  \citenamefont {Wicker}, \citenamefont {Suzuki}, \citenamefont {Jozwiak},
  \citenamefont {Bostwick}, \citenamefont {Rotenberg}, \citenamefont {Bell},
  \citenamefont {Fu}, \citenamefont {Comin},\ and\ \citenamefont
  {Checkelsky}}]{Gapped-nodes-Nature-2018}%
  \BibitemOpen
  \bibfield  {author} {\bibinfo {author} {\bibfnamefont {L.}~\bibnamefont
  {Ye}}, \bibinfo {author} {\bibfnamefont {M.}~\bibnamefont {Kang}}, \bibinfo
  {author} {\bibfnamefont {J.}~\bibnamefont {Liu}}, \bibinfo {author}
  {\bibfnamefont {F.}~\bibnamefont {von Cube}}, \bibinfo {author}
  {\bibfnamefont {C.~R.}\ \bibnamefont {Wicker}}, \bibinfo {author}
  {\bibfnamefont {T.}~\bibnamefont {Suzuki}}, \bibinfo {author} {\bibfnamefont
  {C.}~\bibnamefont {Jozwiak}}, \bibinfo {author} {\bibfnamefont
  {A.}~\bibnamefont {Bostwick}}, \bibinfo {author} {\bibfnamefont
  {E.}~\bibnamefont {Rotenberg}}, \bibinfo {author} {\bibfnamefont {D.~C.}\
  \bibnamefont {Bell}}, \bibinfo {author} {\bibfnamefont {L.}~\bibnamefont
  {Fu}}, \bibinfo {author} {\bibfnamefont {R.}~\bibnamefont {Comin}},\ and\
  \bibinfo {author} {\bibfnamefont {J.~G.}\ \bibnamefont {Checkelsky}},\
  }\bibfield  {title} {\bibinfo {title} {Massive dirac fermions in a
  ferromagnetic kagome metal},\ }\href {https://doi.org/10.1038/nature25987}
  {\bibfield  {journal} {\bibinfo  {journal} {Nature}\ }\textbf {\bibinfo
  {volume} {555}},\ \bibinfo {pages} {638} (\bibinfo {year}
  {2018})}\BibitemShut {NoStop}%
\bibitem [{\citenamefont {Singh}\ \emph {et~al.}(2016)\citenamefont {Singh},
  \citenamefont {Garcia-Castro}, \citenamefont {Valencia-Jaime}, \citenamefont
  {Mu\~noz},\ and\ \citenamefont {Romero}}]{SS_PRB2016}%
  \BibitemOpen
  \bibfield  {author} {\bibinfo {author} {\bibfnamefont {S.}~\bibnamefont
  {Singh}}, \bibinfo {author} {\bibfnamefont {A.~C.}\ \bibnamefont
  {Garcia-Castro}}, \bibinfo {author} {\bibfnamefont {I.}~\bibnamefont
  {Valencia-Jaime}}, \bibinfo {author} {\bibfnamefont {F.}~\bibnamefont
  {Mu\~noz}},\ and\ \bibinfo {author} {\bibfnamefont {A.~H.}\ \bibnamefont
  {Romero}},\ }\bibfield  {title} {\bibinfo {title} {Prediction and control of
  spin polarization in a weyl semimetallic phase of bisb},\ }\href
  {https://doi.org/10.1103/PhysRevB.94.161116} {\bibfield  {journal} {\bibinfo
  {journal} {Phys. Rev. B}\ }\textbf {\bibinfo {volume} {94}},\ \bibinfo
  {pages} {161116} (\bibinfo {year} {2016})}\BibitemShut {NoStop}%
\bibitem [{\citenamefont {Karki}\ \emph {et~al.}(2022)\citenamefont {Karki},
  \citenamefont {Belbase}, \citenamefont {Acharya}, \citenamefont {Singh},\
  and\ \citenamefont {Ghimire}}]{Karki2022}%
  \BibitemOpen
  \bibfield  {author} {\bibinfo {author} {\bibfnamefont {B.}~\bibnamefont
  {Karki}}, \bibinfo {author} {\bibfnamefont {B.~P.}\ \bibnamefont {Belbase}},
  \bibinfo {author} {\bibfnamefont {G.~B.}\ \bibnamefont {Acharya}}, \bibinfo
  {author} {\bibfnamefont {S.}~\bibnamefont {Singh}},\ and\ \bibinfo {author}
  {\bibfnamefont {M.~P.}\ \bibnamefont {Ghimire}},\ }\bibfield  {title}
  {\bibinfo {title} {Pressure-induced creation and annihilation of weyl points
  in
  ${T}_{d}\text{\ensuremath{-}}{\mathrm{mo}}_{0.5}{\mathrm{w}}_{0.5}{\mathrm{te}}_{2}$
  and
  $1{T}^{\ensuremath{''}}\text{\ensuremath{-}}{\mathrm{mo}}_{0.5}{\mathrm{w}}_{0.5}{\mathrm{te}}_{2}$},\
  }\href {https://doi.org/10.1103/PhysRevB.105.125138} {\bibfield  {journal}
  {\bibinfo  {journal} {Phys. Rev. B}\ }\textbf {\bibinfo {volume} {105}},\
  \bibinfo {pages} {125138} (\bibinfo {year} {2022})}\BibitemShut {NoStop}%
\bibitem [{\citenamefont {Muechler}\ \emph {et~al.}(2020)\citenamefont
  {Muechler}, \citenamefont {Liu}, \citenamefont {Gayles}, \citenamefont {Xu},
  \citenamefont {Felser},\ and\ \citenamefont {Sun}}]{PhysRevB.101.115106}%
  \BibitemOpen
  \bibfield  {author} {\bibinfo {author} {\bibfnamefont {L.}~\bibnamefont
  {Muechler}}, \bibinfo {author} {\bibfnamefont {E.}~\bibnamefont {Liu}},
  \bibinfo {author} {\bibfnamefont {J.}~\bibnamefont {Gayles}}, \bibinfo
  {author} {\bibfnamefont {Q.}~\bibnamefont {Xu}}, \bibinfo {author}
  {\bibfnamefont {C.}~\bibnamefont {Felser}},\ and\ \bibinfo {author}
  {\bibfnamefont {Y.}~\bibnamefont {Sun}},\ }\bibfield  {title} {\bibinfo
  {title} {Emerging chiral edge states from the confinement of a magnetic weyl
  semimetal in ${\mathrm{co}}_{3}{\mathrm{sn}}_{2}{\mathrm{s}}_{2}$},\ }\href
  {https://doi.org/10.1103/PhysRevB.101.115106} {\bibfield  {journal} {\bibinfo
   {journal} {Phys. Rev. B}\ }\textbf {\bibinfo {volume} {101}},\ \bibinfo
  {pages} {115106} (\bibinfo {year} {2020})}\BibitemShut {NoStop}%
\bibitem [{\citenamefont {Kubo}(1957)}]{doi:10.1143/JPSJ.12.570}%
  \BibitemOpen
  \bibfield  {author} {\bibinfo {author} {\bibfnamefont {R.}~\bibnamefont
  {Kubo}},\ }\bibfield  {title} {\bibinfo {title} {Statistical-mechanical
  theory of irreversible processes. i. general theory and simple applications
  to magnetic and conduction problems},\ }\href
  {https://doi.org/10.1143/JPSJ.12.570} {\bibfield  {journal} {\bibinfo
  {journal} {Journal of the Physical Society of Japan}\ }\textbf {\bibinfo
  {volume} {12}},\ \bibinfo {pages} {570} (\bibinfo {year} {1957})},\ \Eprint
  {https://arxiv.org/abs/https://doi.org/10.1143/JPSJ.12.570}
  {https://doi.org/10.1143/JPSJ.12.570} \BibitemShut {NoStop}%
\bibitem [{\citenamefont {Cr\'epieux}\ and\ \citenamefont
  {Bruno}(2001)}]{PhysRevB.64.014416}%
  \BibitemOpen
  \bibfield  {author} {\bibinfo {author} {\bibfnamefont {A.}~\bibnamefont
  {Cr\'epieux}}\ and\ \bibinfo {author} {\bibfnamefont {P.}~\bibnamefont
  {Bruno}},\ }\bibfield  {title} {\bibinfo {title} {Theory of the anomalous
  hall effect from the kubo formula and the dirac equation},\ }\href
  {https://doi.org/10.1103/PhysRevB.64.014416} {\bibfield  {journal} {\bibinfo
  {journal} {Phys. Rev. B}\ }\textbf {\bibinfo {volume} {64}},\ \bibinfo
  {pages} {014416} (\bibinfo {year} {2001})}\BibitemShut {NoStop}%
\bibitem [{Note2()}]{Note2}%
  \BibitemOpen
  \bibinfo {note} {The anomalous Hall conductivity component, $\sigma _{xy}$,
  has been computed by following the relationship: \begin {equation}\label
  {eq:ahc} \sigma _{xy}=-\protect \frac {e^2}{\hbar } \DOTSB \sum@ \slimits@
  _n^{occ} \DOTSI \intop \ilimits@ _{BZ} \protect \frac {d^3k}{(2\pi )^3}
  f_n({\protect \bf {k}}) \Omega _{n,xy} (\protect \bf {k}), \end {equation}
  where $\Omega _{xy}(\protect \bf {k})$=$\DOTSB \sum@ \slimits@ _n^{occ}
  f_n(\protect \bf {k})$$\Omega _{n,xy}(\protect \bf {k})$ corresponds to the
  Berry curvature in the $xy$-plane and it is the result of the summation of
  all the occupied $n$-bands and $f_n\protect \bf (k)$ represents the Fermi
  distribution. \protect \leavevmode {\protect \color {black}In this
  calculations, the $\Omega _{n,xy}(\protect \bf {k})$ Berry curvature is
  estimated such as:} \protect \leavevmode {\protect \color {black} \begin
  {equation} \Omega _{n,xy}(\protect \textbf {k}) = -2i\hbar ^2 \DOTSB \sum@
  \slimits@ _{m \protect \neq n} \protect \frac {\left \langle \psi
  _{n,\protect \textbf {k}}\left |v_x\right | \psi _{m,\protect \textbf
  {k}}\right \rangle \left \langle \psi _{m,\protect \textbf {k}}\left
  |v_y\right | \psi _{n,\protect \textbf {k}}\right \rangle }{\left
  [E_m(\protect \textbf {k})-E_n(\protect \textbf {k})\right ]^2} \end
  {equation}} In the latter equation, $\psi _{n,\protect \textbf {k}}$ and
  $v_x$, $v_y$ are the Bloch functions and the velocity operators,
  respectively. In this step, an 320$\times $320$\times $320 $k$-mesh was
  used.}\BibitemShut {Stop}%
\bibitem [{\citenamefont {Chen}\ \emph {et~al.}(2021)\citenamefont {Chen},
  \citenamefont {Le}, \citenamefont {Fu}, \citenamefont {Lin}, \citenamefont
  {Schnelle}, \citenamefont {Sun},\ and\ \citenamefont
  {Felser}}]{PhysRevB.103.144410}%
  \BibitemOpen
  \bibfield  {author} {\bibinfo {author} {\bibfnamefont {D.}~\bibnamefont
  {Chen}}, \bibinfo {author} {\bibfnamefont {C.}~\bibnamefont {Le}}, \bibinfo
  {author} {\bibfnamefont {C.}~\bibnamefont {Fu}}, \bibinfo {author}
  {\bibfnamefont {H.}~\bibnamefont {Lin}}, \bibinfo {author} {\bibfnamefont
  {W.}~\bibnamefont {Schnelle}}, \bibinfo {author} {\bibfnamefont
  {Y.}~\bibnamefont {Sun}},\ and\ \bibinfo {author} {\bibfnamefont
  {C.}~\bibnamefont {Felser}},\ }\bibfield  {title} {\bibinfo {title} {Large
  anomalous hall effect in the kagome ferromagnet
  ${\mathrm{limn}}_{6}{\mathrm{sn}}_{6}$},\ }\href
  {https://doi.org/10.1103/PhysRevB.103.144410} {\bibfield  {journal} {\bibinfo
   {journal} {Phys. Rev. B}\ }\textbf {\bibinfo {volume} {103}},\ \bibinfo
  {pages} {144410} (\bibinfo {year} {2021})}\BibitemShut {NoStop}%
\bibitem [{\citenamefont {Wang}\ \emph {et~al.}(2016)\citenamefont {Wang},
  \citenamefont {Sun}, \citenamefont {Zhang}, \citenamefont {Pang},\ and\
  \citenamefont {Lei}}]{PhysRevB.94.075135}%
  \BibitemOpen
  \bibfield  {author} {\bibinfo {author} {\bibfnamefont {Q.}~\bibnamefont
  {Wang}}, \bibinfo {author} {\bibfnamefont {S.}~\bibnamefont {Sun}}, \bibinfo
  {author} {\bibfnamefont {X.}~\bibnamefont {Zhang}}, \bibinfo {author}
  {\bibfnamefont {F.}~\bibnamefont {Pang}},\ and\ \bibinfo {author}
  {\bibfnamefont {H.}~\bibnamefont {Lei}},\ }\bibfield  {title} {\bibinfo
  {title} {Anomalous hall effect in a ferromagnetic
  ${\mathrm{fe}}_{3}{\mathrm{sn}}_{2}$ single crystal with a geometrically
  frustrated fe bilayer kagome lattice},\ }\href
  {https://doi.org/10.1103/PhysRevB.94.075135} {\bibfield  {journal} {\bibinfo
  {journal} {Phys. Rev. B}\ }\textbf {\bibinfo {volume} {94}},\ \bibinfo
  {pages} {075135} (\bibinfo {year} {2016})}\BibitemShut {NoStop}%
\bibitem [{\citenamefont {Xiao}\ \emph {et~al.}(2010)\citenamefont {Xiao},
  \citenamefont {Chang},\ and\ \citenamefont {Niu}}]{RevModPhys.82.1959}%
  \BibitemOpen
  \bibfield  {author} {\bibinfo {author} {\bibfnamefont {D.}~\bibnamefont
  {Xiao}}, \bibinfo {author} {\bibfnamefont {M.-C.}\ \bibnamefont {Chang}},\
  and\ \bibinfo {author} {\bibfnamefont {Q.}~\bibnamefont {Niu}},\ }\bibfield
  {title} {\bibinfo {title} {Berry phase effects on electronic properties},\
  }\href {https://doi.org/10.1103/RevModPhys.82.1959} {\bibfield  {journal}
  {\bibinfo  {journal} {Rev. Mod. Phys.}\ }\textbf {\bibinfo {volume} {82}},\
  \bibinfo {pages} {1959} (\bibinfo {year} {2010})}\BibitemShut {NoStop}%
\bibitem [{\citenamefont {Jeong}\ \emph {et~al.}(2022)\citenamefont {Jeong},
  \citenamefont {Yang}, \citenamefont {Kim}, \citenamefont {Kim}, \citenamefont
  {Lee},\ and\ \citenamefont {Han}}]{PhysRevB.105.235145}%
  \BibitemOpen
  \bibfield  {author} {\bibinfo {author} {\bibfnamefont {M.~Y.}\ \bibnamefont
  {Jeong}}, \bibinfo {author} {\bibfnamefont {H.-J.}\ \bibnamefont {Yang}},
  \bibinfo {author} {\bibfnamefont {H.~S.}\ \bibnamefont {Kim}}, \bibinfo
  {author} {\bibfnamefont {Y.~B.}\ \bibnamefont {Kim}}, \bibinfo {author}
  {\bibfnamefont {S.}~\bibnamefont {Lee}},\ and\ \bibinfo {author}
  {\bibfnamefont {M.~J.}\ \bibnamefont {Han}},\ }\bibfield  {title} {\bibinfo
  {title} {Crucial role of out-of-plane sb $p$ orbitals in van hove singularity
  formation and electronic correlations in the superconducting kagome metal
  ${\mathrm{csv}}_{3}{\mathrm{sb}}_{5}$},\ }\href
  {https://doi.org/10.1103/PhysRevB.105.235145} {\bibfield  {journal} {\bibinfo
   {journal} {Phys. Rev. B}\ }\textbf {\bibinfo {volume} {105}},\ \bibinfo
  {pages} {235145} (\bibinfo {year} {2022})}\BibitemShut {NoStop}%
\bibitem [{\citenamefont {Di~Sante}\ \emph {et~al.}(2023)\citenamefont
  {Di~Sante}, \citenamefont {Kim}, \citenamefont {Hanke}, \citenamefont
  {Wehling}, \citenamefont {Franchini}, \citenamefont {Thomale},\ and\
  \citenamefont {Sangiovanni}}]{PhysRevResearch.5.L012008}%
  \BibitemOpen
  \bibfield  {author} {\bibinfo {author} {\bibfnamefont {D.}~\bibnamefont
  {Di~Sante}}, \bibinfo {author} {\bibfnamefont {B.}~\bibnamefont {Kim}},
  \bibinfo {author} {\bibfnamefont {W.}~\bibnamefont {Hanke}}, \bibinfo
  {author} {\bibfnamefont {T.}~\bibnamefont {Wehling}}, \bibinfo {author}
  {\bibfnamefont {C.}~\bibnamefont {Franchini}}, \bibinfo {author}
  {\bibfnamefont {R.}~\bibnamefont {Thomale}},\ and\ \bibinfo {author}
  {\bibfnamefont {G.}~\bibnamefont {Sangiovanni}},\ }\bibfield  {title}
  {\bibinfo {title} {Electronic correlations and universal long-range scaling
  in kagome metals},\ }\href
  {https://doi.org/10.1103/PhysRevResearch.5.L012008} {\bibfield  {journal}
  {\bibinfo  {journal} {Phys. Rev. Res.}\ }\textbf {\bibinfo {volume} {5}},\
  \bibinfo {pages} {L012008} (\bibinfo {year} {2023})}\BibitemShut {NoStop}%
\bibitem [{\citenamefont {Wu}\ \emph {et~al.}(2021)\citenamefont {Wu},
  \citenamefont {Schwemmer}, \citenamefont {M\"uller}, \citenamefont
  {Consiglio}, \citenamefont {Sangiovanni}, \citenamefont {Di~Sante},
  \citenamefont {Iqbal}, \citenamefont {Hanke}, \citenamefont {Schnyder},
  \citenamefont {Denner}, \citenamefont {Fischer}, \citenamefont {Neupert},\
  and\ \citenamefont {Thomale}}]{PhysRevLett.127.177001}%
  \BibitemOpen
  \bibfield  {author} {\bibinfo {author} {\bibfnamefont {X.}~\bibnamefont
  {Wu}}, \bibinfo {author} {\bibfnamefont {T.}~\bibnamefont {Schwemmer}},
  \bibinfo {author} {\bibfnamefont {T.}~\bibnamefont {M\"uller}}, \bibinfo
  {author} {\bibfnamefont {A.}~\bibnamefont {Consiglio}}, \bibinfo {author}
  {\bibfnamefont {G.}~\bibnamefont {Sangiovanni}}, \bibinfo {author}
  {\bibfnamefont {D.}~\bibnamefont {Di~Sante}}, \bibinfo {author}
  {\bibfnamefont {Y.}~\bibnamefont {Iqbal}}, \bibinfo {author} {\bibfnamefont
  {W.}~\bibnamefont {Hanke}}, \bibinfo {author} {\bibfnamefont {A.~P.}\
  \bibnamefont {Schnyder}}, \bibinfo {author} {\bibfnamefont {M.~M.}\
  \bibnamefont {Denner}}, \bibinfo {author} {\bibfnamefont {M.~H.}\
  \bibnamefont {Fischer}}, \bibinfo {author} {\bibfnamefont {T.}~\bibnamefont
  {Neupert}},\ and\ \bibinfo {author} {\bibfnamefont {R.}~\bibnamefont
  {Thomale}},\ }\bibfield  {title} {\bibinfo {title} {Nature of unconventional
  pairing in the kagome superconductors $a{\mathrm{v}}_{3}{\mathrm{sb}}_{5}$
  ($a=\mathrm{K},\mathrm{Rb},\mathrm{Cs}$)},\ }\href
  {https://doi.org/10.1103/PhysRevLett.127.177001} {\bibfield  {journal}
  {\bibinfo  {journal} {Phys. Rev. Lett.}\ }\textbf {\bibinfo {volume} {127}},\
  \bibinfo {pages} {177001} (\bibinfo {year} {2021})}\BibitemShut {NoStop}%
\bibitem [{\citenamefont {Sun}\ \emph {et~al.}(2015)\citenamefont {Sun},
  \citenamefont {Ruzsinszky},\ and\ \citenamefont
  {Perdew}}]{PhysRevLett.115.036402}%
  \BibitemOpen
  \bibfield  {author} {\bibinfo {author} {\bibfnamefont {J.}~\bibnamefont
  {Sun}}, \bibinfo {author} {\bibfnamefont {A.}~\bibnamefont {Ruzsinszky}},\
  and\ \bibinfo {author} {\bibfnamefont {J.~P.}\ \bibnamefont {Perdew}},\
  }\bibfield  {title} {\bibinfo {title} {Strongly constrained and appropriately
  normed semilocal density functional},\ }\href
  {https://doi.org/10.1103/PhysRevLett.115.036402} {\bibfield  {journal}
  {\bibinfo  {journal} {Phys. Rev. Lett.}\ }\textbf {\bibinfo {volume} {115}},\
  \bibinfo {pages} {036402} (\bibinfo {year} {2015})}\BibitemShut {NoStop}%
\bibitem [{\citenamefont {Furness}\ \emph {et~al.}(2020)\citenamefont
  {Furness}, \citenamefont {Kaplan}, \citenamefont {Ning}, \citenamefont
  {Perdew},\ and\ \citenamefont {Sun}}]{doi:10.1021/acs.jpclett.0c02405}%
  \BibitemOpen
  \bibfield  {author} {\bibinfo {author} {\bibfnamefont {J.~W.}\ \bibnamefont
  {Furness}}, \bibinfo {author} {\bibfnamefont {A.~D.}\ \bibnamefont {Kaplan}},
  \bibinfo {author} {\bibfnamefont {J.}~\bibnamefont {Ning}}, \bibinfo {author}
  {\bibfnamefont {J.~P.}\ \bibnamefont {Perdew}},\ and\ \bibinfo {author}
  {\bibfnamefont {J.}~\bibnamefont {Sun}},\ }\bibfield  {title} {\bibinfo
  {title} {Accurate and numerically efficient r2scan meta-generalized gradient
  approximation},\ }\href {https://doi.org/10.1021/acs.jpclett.0c02405}
  {\bibfield  {journal} {\bibinfo  {journal} {The Journal of Physical Chemistry
  Letters}\ }\textbf {\bibinfo {volume} {11}},\ \bibinfo {pages} {8208}
  (\bibinfo {year} {2020})},\ \bibinfo {note} {pMID: 32876454},\ \Eprint
  {https://arxiv.org/abs/https://doi.org/10.1021/acs.jpclett.0c02405}
  {https://doi.org/10.1021/acs.jpclett.0c02405} \BibitemShut {NoStop}%
\end{thebibliography}%

\end{document}